\documentclass[]{spie}

\usepackage{amsmath,amsfonts,amssymb}
\usepackage{graphicx}
\usepackage[colorlinks=true, allcolors=blue]{hyperref}

\title{Highly-multiplexed microwave SQUID readout using the SLAC Microresonator Radio Frequency (SMuRF) Electronics for Future CMB and Sub-millimeter Surveys}

\author[a,b]{Shawn W. Henderson}
\author[a,b]{Zeeshan Ahmed}
\author[c]{Jason Austermann}
\author[d]{Daniel Becker}
\author[c]{Douglas A. Bennett}
\author[b]{David Brown}
\author[e]{Saptarshi Chaudhuri}
\author[a]{Hsiao-Mei Sherry Cho}
\author[b]{John M. D'Ewart}
\author[c]{Bradley Dober}
\author[c]{Shannon M. Duff}
\author[b]{John E. Dusatko}
\author[f]{Sofia Fatigoni}
\author[b]{Josef C. Frisch}
\author[d]{Jonathon D. Gard}
\author[f]{Mark Halpern}
\author[c]{Gene C. Hilton}
\author[c]{Johannes Hubmayr}
\author[a,b,c]{Kent D. Irwin}
\author[e]{Ethan D. Karpel}
\author[e]{Sarah S. Kernasovskiy}
\author[e]{Stephen E. Kuenstner}
\author[a,b,c]{Chao-Lin Kuo}
\author[a]{Dale Li}
\author[d]{John A. B. Mates}
\author[c]{Carl D. Reintsema}
\author[b]{Stephen R. Smith}
\author[c,d]{Joel Ullom}
\author[c]{Leila R. Vale}
\author[b]{Daniel D. Van Winkle}
\author[c]{Michael Vissers}
\author[e]{Cyndia Yu}

\affil[a]{Kavli Institute for Particle Astrophysics and Cosmology, Menlo Park, CA 94025, USA}
\affil[b]{SLAC National Accelerator Laboratory, Menlo Park, CA 94025, USA}
\affil[c]{National Institute of Standards and Technology, Boulder, CO, USA}
\affil[d]{University of Colorado Boulder, Boulder, CO 80309}
\affil[e]{Department of Physics, Stanford University, Stanford, CA 94305, USA}
\affil[f]{Department of Physics and Astronomy, University of British Columbia, Vancouver, BC V6T 1Z4, Canada}

\authorinfo{Corresponding Author: S. W. H.  Email: shawn at slac dot stanford dot edu}

\begin{document} 

\maketitle
 
\begin{abstract} 
The next generation of cryogenic CMB and submillimeter cameras under development require densely instrumented sensor arrays to meet their science goals. The readout of large numbers ($\sim$10,000--100,000 per camera) of sub-Kelvin sensors, for instance as proposed for the CMB-S4 experiment, will require substantial improvements in cold and warm readout techniques. To reduce the readout cost per sensor and integration complexity, efforts are presently focused on achieving higher multiplexing density while maintaining readout noise subdominant to intrinsic detector noise and presenting manageable thermal loads. Highly-multiplexed cold readout technologies in active development include Microwave Kinetic Inductance Sensors (MKIDs) and microwave rf-SQUIDs. Both exploit the high quality factors of superconducting microwave resonators to densely channelize sub-Kelvin sensors into the bandwidth of a microwave transmission line. In the case of microwave SQUID multiplexing, arrays of transition-edge sensors (TES) are multiplexed by coupling each TES to its own superconducting microwave resonator through an rf-SQUID. We present advancements in the development of a new warm readout system for microwave SQUID multiplexing, the SLAC Superconducting Microresonator RF electronics, or SMuRF, which is built on the SLAC National Accelerator Laboratory's Advanced Telecommunications Computing Architecture (ATCA) FPGA Common Platform. SMuRF aims to read out 4000 microwave SQUID channels between 4 and 8 GHz per RF line. Each compact SMuRF system is built onto a single ATCA carrier blade. Daughter boards on the blade implement RF frequency-division multiplexing using FPGAs, fast DACs and ADCs, and an analog up- and down-conversion chain. The system reads out changes in flux in each resonator-coupled rf-SQUID by monitoring the change in the transmitted amplitude and frequency of RF tones produced at each resonator's fundamental frequency. The SMuRF system is unique in its ability to track each tone, minimizing the total RF power required to read out each resonator, thereby significantly reducing the linearity requirements on the cold and warm readout. Here, we present measurements of the readout noise and linearity of the first full SMuRF system, including a demonstration of closed-loop tone tracking on a 528 channel cryogenic microwave SQUID multiplexer. SMuRF is being explored as a potential readout solution for a number of future CMB projects including Simons Observatory, BICEP Array, CCAT-prime, Ali-CPT, and CMB-S4. In addition, parallel development of the platform is underway to adapt SMuRF to read out both MKID and fast X-ray TES calorimeter arrays.
\end{abstract}

\keywords{FPGA-based RF Readout Electronics, Multiplexing, Microwave SQUID Multiplexing, Tone tracking, Linearity, Transition Edge Sensor Arrays, MKIDs, Cosmic Microwave Background (CMB)}

\section{Introduction}

Planned next generation surveys seek to map the mm- and sub-mm sky with unprecedented sensitivity using ultra-sensitive large format superconducting sensor arrays.
These future surveys have the potential to radically transform our understanding of fundamental physics and astronomy, from the physics of inflation to details of the formation of the first stars and galaxies~\cite{CMBs4Science}.
While a variety of superconducting sensor array technologies are under development~\cite{CMBs4Tech}, on-going Stage-3 CMB experiments including Advanced ACT~\cite{AdvACT16}, BICEP Array~\cite{Grayson16}, POLARBEAR-2~\cite{Suzuki16} and SPT-3G~\cite{Anderson2018} are presently fielding 150-mm transition edge sensor (TES) arrays with thousands of TESs~\cite{Duff16,Posada15,Suzuki12}.
TES arrays are the most mature superconducting sensor technology, achieve photon-noise-limited performance across the full range of optical frequencies (30--300~GHz) of interest for future ground-based mm- and sub-mm surveys, and have a long heritage of on-sky performance~\cite{CMBs4Tech}.

However, the readout of TES arrays is currently limited by low-density signal multiplexing techniques.~\cite{CMBs4Tech} 
Time-division multiplexing using DC-SQUIDs (TDM) and frequency-division multiplexing using MHz LC resonators and DC-SQUIDs (DfMux) require thousands of wires from room temperature to cryogenic temperatures because of multiplexing (MUX)\footnote{Multiplexing (MUX) factor is the number of sensor signals utilizing a single readout chain of wires from cryogenic to room temperatures.} factors of 64 and 68 respectively.~\cite{Henderson16, Bender14}.
They also require a large number of superconducting interconnects between SQUIDs and other components at cryogenic temperatures and the TES arrays. 
Increasing the multiplexing density by more than a factor of two to reduce system complexity degrades noise performance for these readout techniques.
Thus, these techniques are likely too expensive, or cumbersome, or provide insufficient noise performance for the next generation of large receivers, such as those planned for CMB-S4, which aspires to accommodate as many as $100,000$ TES sensors per receiver.

Microwave frequency-division multiplexing (FDM) techniques address readout scaling challenges by densely packing the signals from potentially thousands of superconducting sensors within the bandwidth of one common high bandwidth microwave feedline using superconducting microresonators~\cite{Zmuidzinas12}.
FDM techniques utilizing GHz resonators under development include microwave kinetic inductance detectors (MKIDs)~\cite{Day03} and microwave SQUID ($\mu$MUX) readout of TESs~\cite{Irwin2004,Lehnert07,Mates11PhD}. 
Both techniques share a common basis of GHz excitation and readout techniques. 
In MKID readout, the superconducting sensors are not TESs but the GHz resonators themselves and the read out of up to 1000 MKIDs on a single coaxial line has been demonstrated in the lab~\cite{vanRantwijk16}.
In $\mu$MUX readout, each TES is inductively coupled through an rf-SQUID to its own resonator at a unique microwave frequency.
Multiplexing factors exceeding $1000$ are possible to achieve for the read out of existing, low-noise TES arrays without degrading noise performance. 
Previous implementations of $\mu$MUX have achieved multiplexing factors of $128$~\cite{Mates17}, and a 64 MUX factor system was deployed on the Green Bank Telescope for MUSTANG-2~\cite{Stanchfield16}. 
These implementations have utilized cold multiplexers limited by resonator bandwidth and spacing density, and variants of a warm readout originally developed for MKIDs based on the CASPER ROACH-2 platform~\cite{McHugh12,Duan10,Hickish16,Madden17,Gard2018}, which are limited by system linearity.
The constraints on cold resonator design have been obviated, as will be addressed in a future publication~\cite{DoberSPIE18}. The warm readout system linearity limitations have been solved as well and are addressed here. 

In this paper, Section~\ref{sec:tonetracking} describes the system linearity challenges to simultaneously read out more than ${\cal O}(100)$ TES sensors and how the SLAC Microresonator RF Electronics (SMuRF) solve this problem.
In Section~\ref{sec:smurfoverview} we describe the SMuRF system.
Section~\ref{sec:smurfperformance} presents a first validation of SMuRF performance on a 528-channel $\mu$MUX multiplexer fabricated at the National Institute of Standards and Technology (NIST) in Boulder.
Reading out this 528-channel $\mu$MUX cryogenic multiplexer the SMuRF readout achieves the highest $\mu$MUX multiplexing factor yet demonstrated, simultaneously reading out $426$ channels, with negligible observed degradation in readout noise.

\section{Linearity and tone tracking}
\label{sec:tonetracking}

Existing $\mu$MUX and MKID readout systems read out superconducting resonators using fixed frequency RF probe tones generated at frequencies on or near the resonators' characteristic frequencies.
For an RF system generating tones in an octave of bandwidth, nonlinearities in every active component in the signal path will generate unwanted 3rd order intermodulation products that will fall directly in-band. 
These unwanted parasitic in-band tones will interfere with and distort the fundamental RF probe tones.  
The theoretical number of 3rd order tones generated in an $N$ tone system is $2N(N-1)$.  
In a readout generating 2000 tones, nearly 8~million 3rd order intermodulation product tones will be produced. 
The power in these 3rd order intermodulation products grows as the cube of the power in the fundamental tones~\cite{Pozar11}. 
For large multiplexing factors and sufficiently high power, the intermodulation products become an irreducible pseudo-noise floor.

Warm readout systems with fixed frequency RF probe tones  provide insufficient linearity to read out $\mu$MUX cryogenic multiplexers with MUX factors exceeding a mere $\cal{O}$(100) resonators.
This is because optimal $\mu$MUX probe tone powers are 20--30~dB higher than typical MKID probe tone powers of $-100$~dBm.
Relative to MKIDs, the higher required power per tone in $\mu$MUX significantly relaxes the requirements on the noise temperature of the cryogenic amplifier used to amplify the tones for read out, but the larger summed power presents a linearity challenge for commercially available electronics, particularly cryogenic amplifiers.
Additionally, $\mu$MUX resonator frequencies are constantly shifting as their rf-SQUIDs are flux ramp modulated~\cite{Mates12}.
Flux ramp modulation is a scheme for linearizing the rf-SQUID response without running feedback wires to each SQUID. 
Beyond linearizing the SQUID responses, flux ramp modulation enables $\mu$MUX systems to overcome two-level system (TLS) noise.
However, for readout systems which use fixed tones to probe the $\mu$MUX resonances, the transmitted tone power can vary by 10--20~dB per tone as the resonances are flux ramp modulated, continuously varying the total power at the input of the cold RF chain~\cite{Kernasovskiy18}.

SMuRF solves the $\mu$MUX linearity problem through its incorporation of the latest in high linearity RF components, including a two-stage cryogenic amplifier chain optimized for linearity, and a novel FPGA implementation of closed-loop tone tracking.
Section~\ref{subsec:rfamc} presents measurements of the linearity of the warm SMuRF electronics in isolation and Section~\ref{subsec:cryostat} presents a two-stage cryogenic amplifier chain with substantially improved linearity over standard implementations of $\mu$MUX readout~\cite{Giachero16}. 
Section~\ref{subsec:trackingalgo} details the algorithm used to simultaneously maintain tones on resonance while they are flux ramp modulated, at all times minimizing the summed power transmitted into the subsequent RF chain.
This paper builds on a prior 64 channel tone tracking demonstration with a SMuRF prototype system~\cite{Kernasovskiy18}.

\section{System overview}
\label{sec:smurfoverview}

\begin{figure}[hbt!]
  \centering
  \includegraphics[width=\textwidth]{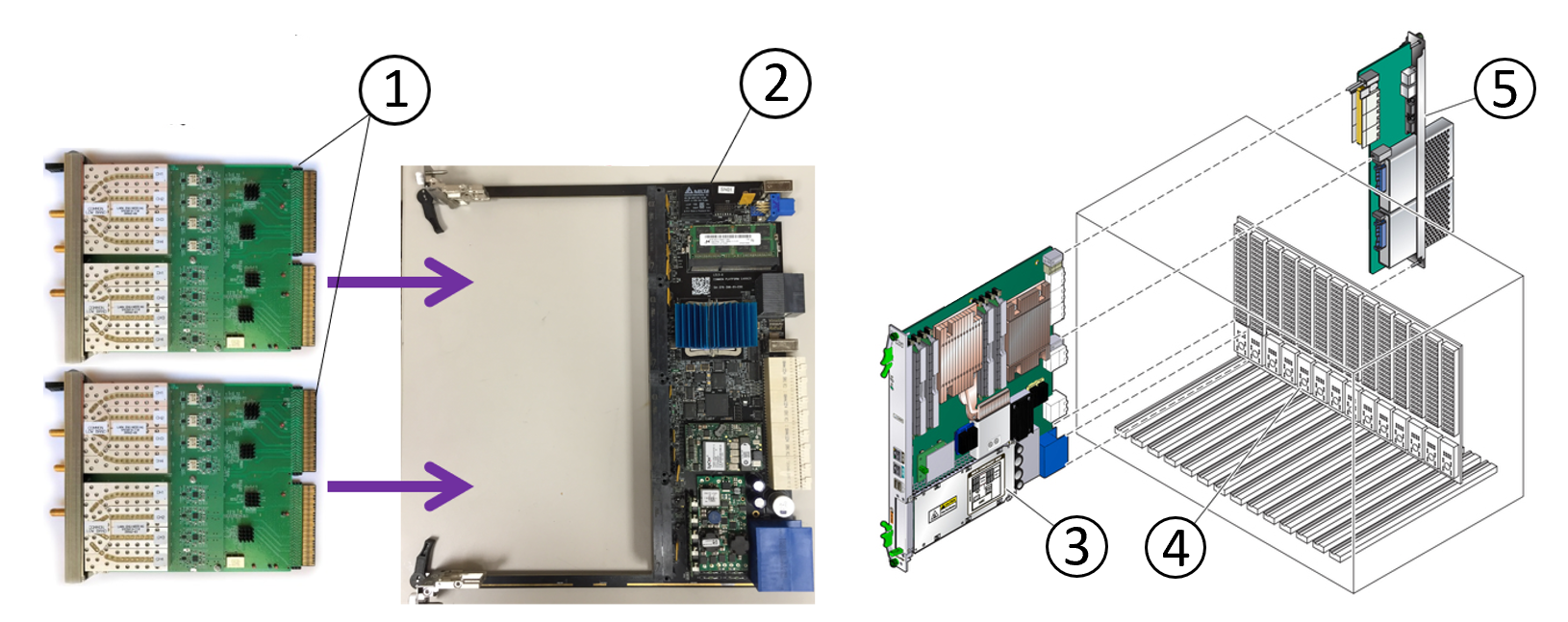}
  \caption{A complete SMuRF readout system and its components.  As labeled, {\it 1)} High and low band RF Advanced Mezzanine Cards (AMCs) described in Section~\ref{subsec:rfamc}. {\it 2)} FPGA carrier card described in Section~\ref{subsec:carrier}. {\it 3)}  A complete SMuRF readout card consists of an FPGA carrier card integrated with one low band and one high band RF AMC card. {\it 4)} The SMuRF cards are housed in a commercial Advanced Telecommunications Computing Architecture (ATCA) crate. {\it 5)}  Each SMuRF readout card additionally interfaces through the ATCA crate backplane with a Rear Transition Module (RTM).  The RTM is described in Section~\ref{subsec:rtmandcryocard}. {\it Not shown:} The RTM is connected to a cryostat card through a shielded twisted pair cable.  The cryostat card interfaces signals from the RTM with an external cryostat housing the cold RF circuit and $\mu$MUX multiplexer.}
  \label{fig:smurfsystem}
\end{figure}

In this section we briefly describe the most critical components of the SMuRF system shown in Figure~\ref{fig:smurfsystem}.  A more complete description of the SMuRF system components, their design, and their optimization will be presented in a future publication.

\subsection{SLAC Common Platform carrier card}
\label{subsec:carrier}
The carrier card, built on the SLAC National Accelerator Laboratory's Advanced Telecommunications Computing Architecture (ATCA) FPGA Common Platform, serves as the backbone for the SMuRF readout system.
The carrier card contains an FPGA which provides data communication to the Advanced Mezzanine Cards (AMCs) and a Rear Transition Module (RTM) card, runs the signal processing and feedback algorithms, and provides an ethernet interface to the data acquisition system~\cite{Till18}.   
It has two card bays, for double-wide, full-height AMC cards, and an interface to an RTM.

SMuRF currently uses a carrier card with a Xilinx KU060 Kintex Ultrascale FPGA, and will be upgraded to a KU15P Ultrascale+ part in the near future.  It provides 8X, 12.5Gb/s uplink and downlink JESD204b interfaces to each AMC card, as well as a 10Gb/s ethernet link to the ATCA crate backplane. An additional ethernet link is used to receive serial timing data to synchronize all of the SMuRF cards in the system. 

The carrier card includes 8GB of DDR3 memory for recording long diagnostic data buffers, and 3MB of fast on-FPGA cache capable of operating at full FPGA speeds. IPMI management and power switching is supported according to the ATCA specification~\cite{ATCAspec}.  

\subsection{RF Advanced Mezzanine Cards}
\label{subsec:rfamc}
In addition to generating the RF tones, the RF Advanced Mezzanine Cards (AMCs) serve the function of digitizing the input RF signals and converting them into 500 MHz blocks to be processed by the digital signal processing code in the FPGA contained on the carrier card. 
There are two types of RF AMC cards, the high (6-8~GHz) and low (4-6~GHz) band cards.   Each of the AMC cards is assembled from a base card (called the SMuRF base card) as well as a daughter card (called the high or low band daughter card) which contains most of the high frequency electronics (with the exception of the LO generation).

The SMuRF base card contains all the necessary circuitry to generate the 4 LO (local oscillator tones) necessary for down/up conversion as well as 2 ADC (ADC32RF44) and 2 DAC (DAC38J84) chips, each of which contain 2 ADCs and 2 DACs respectively for a total of 4 ADCs and 4 DACs each used to generate and detect the 500 MHz bands.  In addition, the board has a clock generation chip (LMK 04828) used to generate all the clockwork and synchronization for the high speed JESD204B lanes.  Finally, there is a clean VCXO on board to serve as either an independent or lockable reference for the system.  

The high and low band cards are functionally equivalent with the main difference being the frequency range of the high frequency filter/combiner (quadruplexer) used to combine (from the up-converters) or split signals (before the down-converters).  In addition, the high band card contains a broad-band combiner for the RF output to combine the high and low band RF signals for an overall bandwidth of 4-8 GHz as well as a splitter for splitting off the low band signal to send to the low band card from the RF input.  Each card contains 4 up-converters and 4 down converters.  All individual channels have programmable step attenuators used for leveling the 500 MHz bands.  

\begin{figure*}
\begin{center}
\resizebox{\textwidth}{!}{
\includegraphics[width=0.500\textwidth]{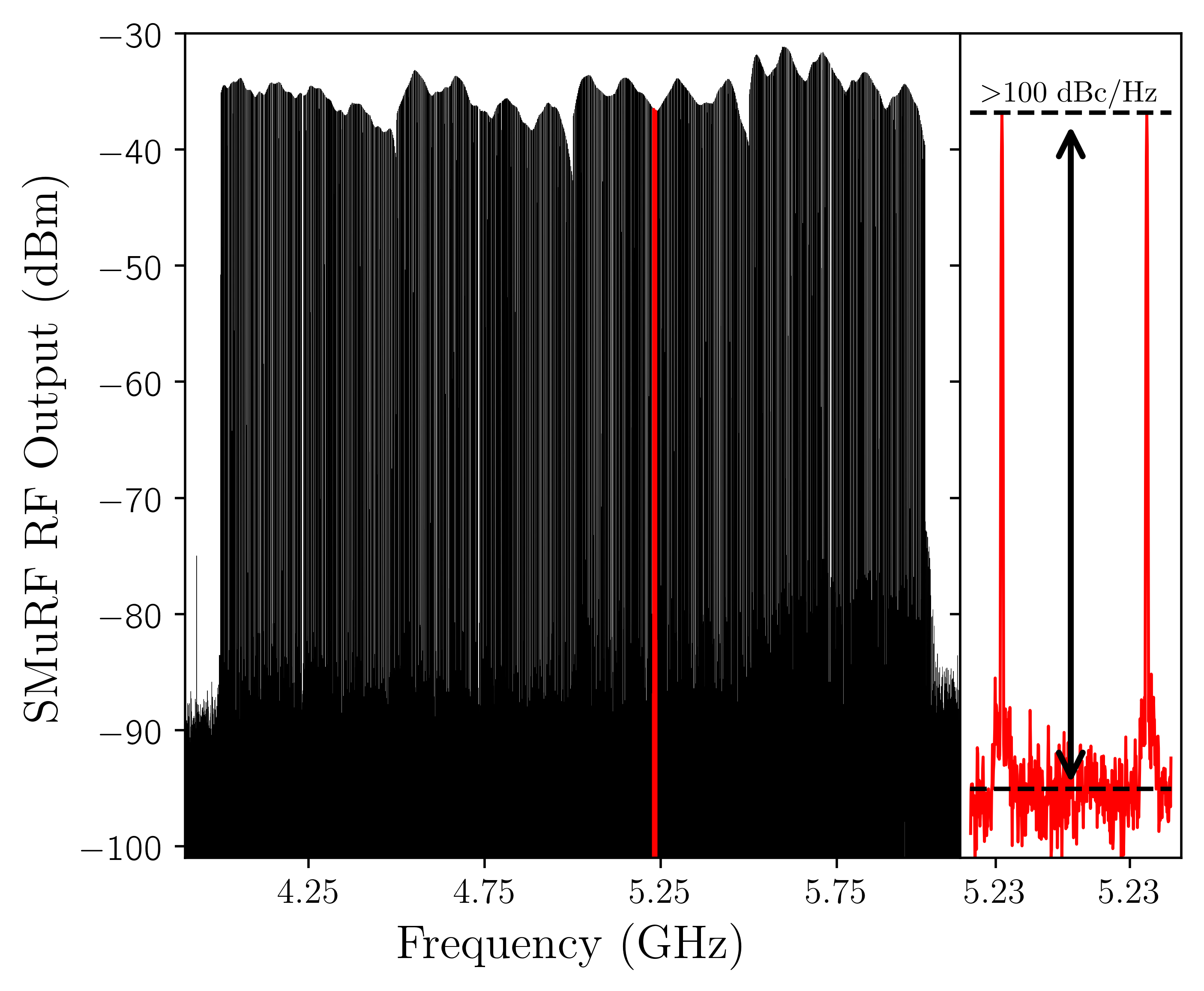}
\includegraphics[width=0.500\textwidth]{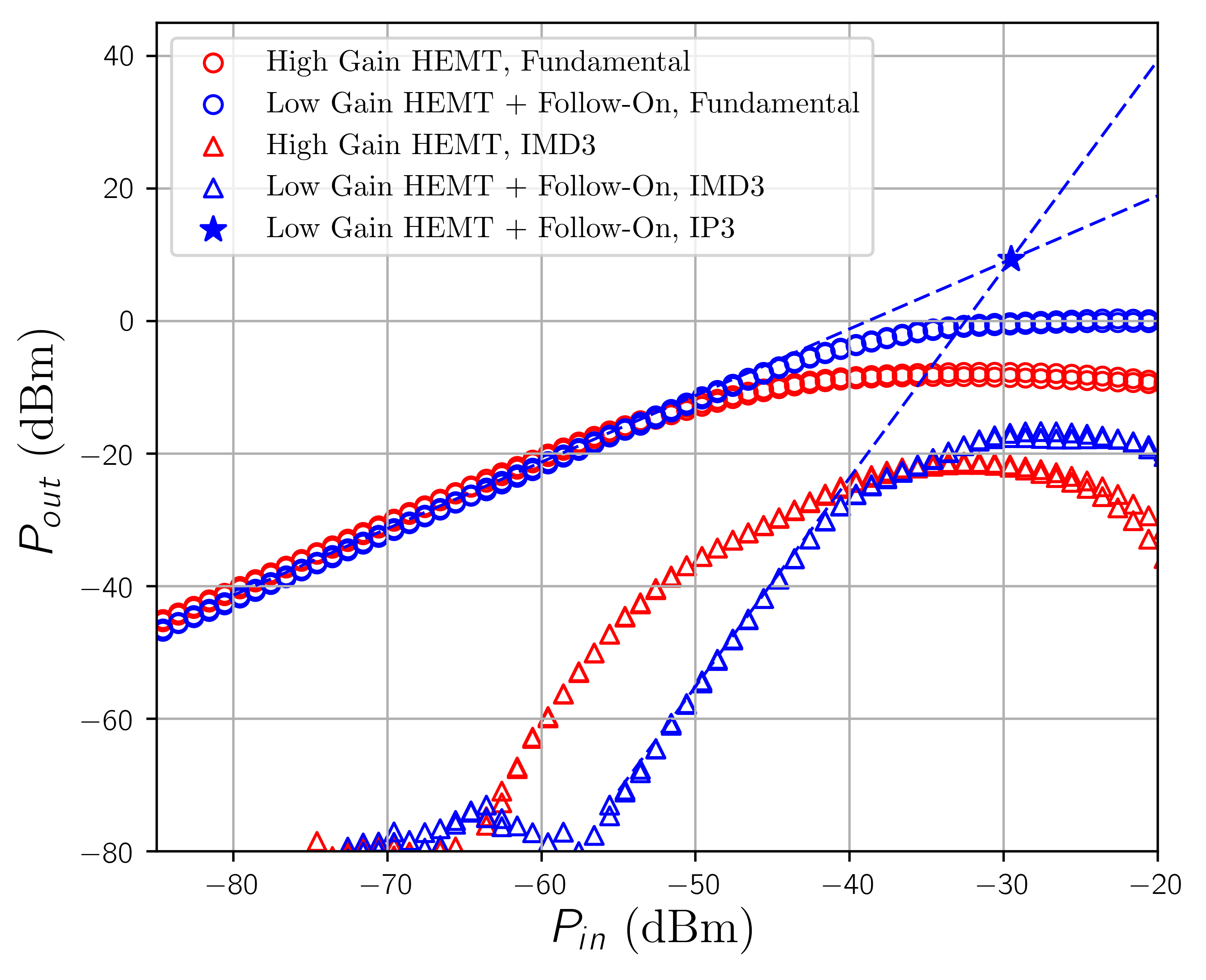}}
\end{center}
\caption{{\it Left:} 1000 lines generated between 4--6~GHz and read back directly through a short cable with a SMuRF low band RF AMC card.  The SMuRF warm readout alone achieves a dynamic range exceeding $100$~dBc/Hz even when generating 1000 tones in a 2~GHz bandwidth as shown by the data highlighted in red.  Note that the data in this plot has not been corrected for the resolution bandwidth of the signal analyzer, resulting in an apparent signal to noise (reading the y-axis) much less than $100$~dBc/Hz.
{\it Right:} Measured 1~dB compression point and third-order intercept point (IP3 -- the blue star) for the high linearity two-stage cold amplifier chain developed to exploit the linearity of the SMuRF electronics. 
} 
\label{fig:linearity}
\end{figure*}

Figure~\ref{fig:linearity} highlights measurements of the linearity of the SMuRF system. 
To measure the SMuRF output signal to noise we used the DAC/Upconverter section on a low band RF AMC to generate 1000 lines with frequencies distributed between 4 and 6 GHz.  
The frequency spacing between each line was randomly drawn from a gaussian distribution with a standard deviation of 1.8~MHz.
An intentional gap was left in the spectrum to allow for noise floor measurements.  
The measured signal to noise for output tones measured using a spectrum analyzer exceeds 100 dBc/Hz across the full 4--6~GHz bandwidth.  
Further, we looped the $1000$ DAC-generated tones back into the SMuRF input (which typically receives signals from the cryostat) in order to measure the linearity of the full SMuRF system end-to-end.  
In this loopback mode, the measured signal to noise also exceeded 100~dBc/Hz over the full 4--6~GHz bandwidth of the low band RF AMC.

\subsection{Rear transition module and cryostat card}
\label{subsec:rtmandcryocard}
The RTM board generates TES differential voltage biases via 32 on-board serial DACs.  In addition, a 50~MHz parallel DAC is used to generate the flux ramp.  Special care has been taken to ensure that all DACS, buffers, voltage references, resistors, and capacitors meet stringent low frequency and temperature stability requirements.  A smaller card, which is located on the cryostat, connects to the RTM.  This cryostat card contains temperature stable bias resistors and filters for the TES voltage bias lines as well as relays used to delatch superconducting TESs by switching to lower bias resistors for higher currents.
In addition, a microcontroller can be used to control and servo the bias currents or voltages required by cryogenic amplifiers.

The RTM card connects to the carrier card via a dedicated RTM communication connector.  In addition, 12V power is supplied from the carrier via a separate ATCA standard RTM connector.  Communication to both the RTM and cryostat cards is done via standard Serial Peripheral Interface (SPI) busses to the main FPGA on the carrier card.  SPI is used to communicate between the main FPGA on the carrier card and a CPLD on the RTM.  SPI is also used for communication between the main FPGA on the carrier card and a microcontroller on the cryostat card.

Both the RTM and cryostat card are designed to be operated in static mode.  In static mode the TES biases are set after which nearly all logic operations on the boards cease.  An exception on the RTM is the flux ramp signal, which during normal operation is clocked at approximately 50~MHz.  The microcontroller on the cryostat board is not expected to be clocking except when woken up for TES delatching or periodically monitoring the bias of cryogenic amplifiers.

\subsection{Firmware}

SMuRF is based on the SLAC controls common platform~\cite{Frisch17}.  The top level firmware is partitioned into two sections - a common platform core and an application core.  The common platform core is based on the open-source SLAC Ultimate RTL Framework (SURF)~\footnote{https://github.com/slaclab/surf}.  SURF is a VHDL library which contains a library of protocols, device access, and commonly used modules.  SURF includes Ethernet IP, AXI4 memories interface, synchronization IP, serial protocols, commonly used device interfaces, and wrapped Xilinx IPs.

SMuRF digital signal processing is contained in the application core.  SMuRF digitizes and processes up to 4000 channels in a 4~GHz bandwidth.  The 4~GHz bandwidth is first split into 500~MHz blocks by the analog front end electronics and shifted to 750~MHz centered intermediate frequency (IF).  RF ADCs oversample the IF and perform digital down conversion to complex baseband with a data rate of 614.4~MHz.  SMuRF firmware interfaces to the ADC stream via the SURF JESD204B IP core.   

An analysis filter bank performs coarse first stage down conversion and frequency-to-time multiplex conversion.  The analysis filter bank splits the 614.4~MHz complex baseband data into 128 overlapping 9.6~MHz wide sub-bands, each with 5~MHz usable bandwidth. The time multiplexed sub-bands go through a second stage downconversion to baseband.  Presently the SMuRF firmware is able to downconvert up to four tones per sub-band for a total of 512 tones in 614.4~MHz band.  The SMuRF baseband processor performs tone tracking and flux ramp demodulation.  The tracking algorithm tunes a set of numeric controlled oscillators (NCOs) such that the generated tone interrogates the resonator on resonance.  Next, the time multiplexed NCO outputs go through a synthesis filter bank, combining the 128 sub-bands, for time-to-frequency multiplexing conversion.  The complex data stream is sent in complex baseband to an RF DAC which performs a final digital up conversion (DUC) to 750~MHz IF.

\subsection{Tracking algorithm and demodulation}
\label{subsec:trackingalgo}

SMuRF tracks resonator frequency shifts and maintains each RF probe tone on resonance using a closed feedback loop operating independently on each tone. 
For each subsequent iteration of the feedback loop, SMuRF estimates the shift in each resonator's frequency $\Delta f$ using a prior measurement of the complex transmission near each resonance.
For frequencies near resonance the quadrature $Q$ and in-phase $I$ components of a transmitted probe tone trace a circle in the complex $(I,Q)$ plane.
The complex response for each resonator is rotated and scaled into new complex coordinates $(I^{\prime},Q^{\prime})$ such that any resonator frequency variation contributes only in the transformed quadrature direction $Q^{\prime}$~\cite{Gao07}.
The measurement of each resonator's complex response must only be performed once before initiating tracking as described in Section~\ref{subsec:tuning}.
Perfectly on resonance, $Q^{\prime}=0$, and
$Q^{\prime}$ is used as an error term in the tone tracking algorithm, where feedback seeks to drive $Q^{\prime}$ to zero by varying the frequency of the output probe tone.
The tracking algorithm relies on the measured resonator response to compute the equivalent change in resonator frequency $\Delta f$. 
SMuRF keeps each probe tone on resonance by adjusting its frequency to null the measured tracked frequency error $\Delta f$. 

The change in resonator frequency $\Delta f$ with time due to the flux-ramp-modulated SQUID response is expected to be periodic and of the approximate form 
\begin{equation}
\label{eqn:freqshift}
\Delta f(t)=C\Bigg(\frac{\lambda\cos(w_{c}t + \phi(t))}{1+\lambda\cos(w_{c}t + \phi(t))}\Bigg),
\end{equation}
where $C$ and $\lambda$ are constants depending on the design and material properties of the resonators and rf-SQUIDs~\cite{Mates11PhD}.  The carrier frequency $f_{c} = \omega_{c}/2\pi$ is determined by the product of the flux ramp reset rate (typically 10-30~kHz) and the number of flux quanta swept over each flux ramp period in the rf-SQUIDs (typically 3-6).
Typical design values for the resonators in the 528-channel multiplexer tested in these proceedings are $C\sim100$~kHz and $\lambda\sim0.3$ (unit-less)~\cite{DoberSPIE18}.  See Figure~\ref{fig:tuning} for a typical measured resonator $\Delta f(t)$ due to flux ramp modulation.

The resonator frequency as a function of time $f(t)$ can be expressed in terms of a Fourier series
\begin{equation}
\label{eqn:fourier_series}
f(t) = a_{0} + \sum_{n=1}^{\infty} a_{n} \cos(n w_{c} t) + b_{n} \sin(n w_{c}t)
\end{equation}

The Fourier series representation provides a linear model for the unknown Fourier coefficients.    
We approximate the SQUID response to $n=3$, or the first three harmonics, and use feedback to track $a_{0}$, $a_{n}$, $b_{n}$ for ${n=1...3}$.  
The phase change of each harmonic is proportional to a phase change $\Delta\phi$ from TES current ( $\Delta \arctan({b_{n}}/{a_{n}}) = n\Delta\phi$ ).  
This allows the tracking algorithm to also perform demodulation via an $\arctan$ on the tracked parameters.  
The signal in each harmonic varies with each SQUID and with probe tone power, but does not need to be known {\it a priori}. 
Further details on SMuRF's tracking algorithm and its performance will be presented in a future publication.

\subsection{Software}
\label{subsec:software}
FPGA communication goes over a 10Gbps Ethernet link.  The Reliable SLAC Streaming Interface (RSSI) provides reliable in-order delivery of UDP packets~\cite{RSSI}. 

Rogue, a C++ library with Python bindings, is used for Low-level FPGA communications (register access, DDR interface, asynchronous streaming).  The Python interface (PyRogue~\footnote{https://github.com/slaclab/pyrogue-control-server}) is used for device description (the register map) and allows for rapid development.
Rogue also includes an EPICS channel access server plugin.  This allows a simple register get/put interface from a variety of high level clients.  Both MATLAB and Python implementations of high level control applications are under active development.  

\section{System performance}
\label{sec:smurfperformance}

In this section we present first results from testing an integrated SMuRF system on a 528 channel NIST microwave SQUID multiplexer, described in Section~\ref{subsec:NIST528}.  
The multiplexer was installed on the 250~mK temperature stage of a $^{3}$He sorption fridge, as described in Section~\ref{subsec:cryostat}.  
The SMuRF FPGA carrier card for these tests was populated with only one low band RF AMC running tone tracking digital signal processing implemented over a 1~GHz bandwidth.
Details on the system configuration and operation are given in Section~\ref{subsec:tuning} and Section~\ref{subsec:noise} presents first noise measurements.

\subsection{NIST 528-channel multiplexer}
\label{subsec:NIST528}
\begin{figure}[!hbt]
	\centering
	\label{fig:528vna}
	\includegraphics[width=\textwidth]{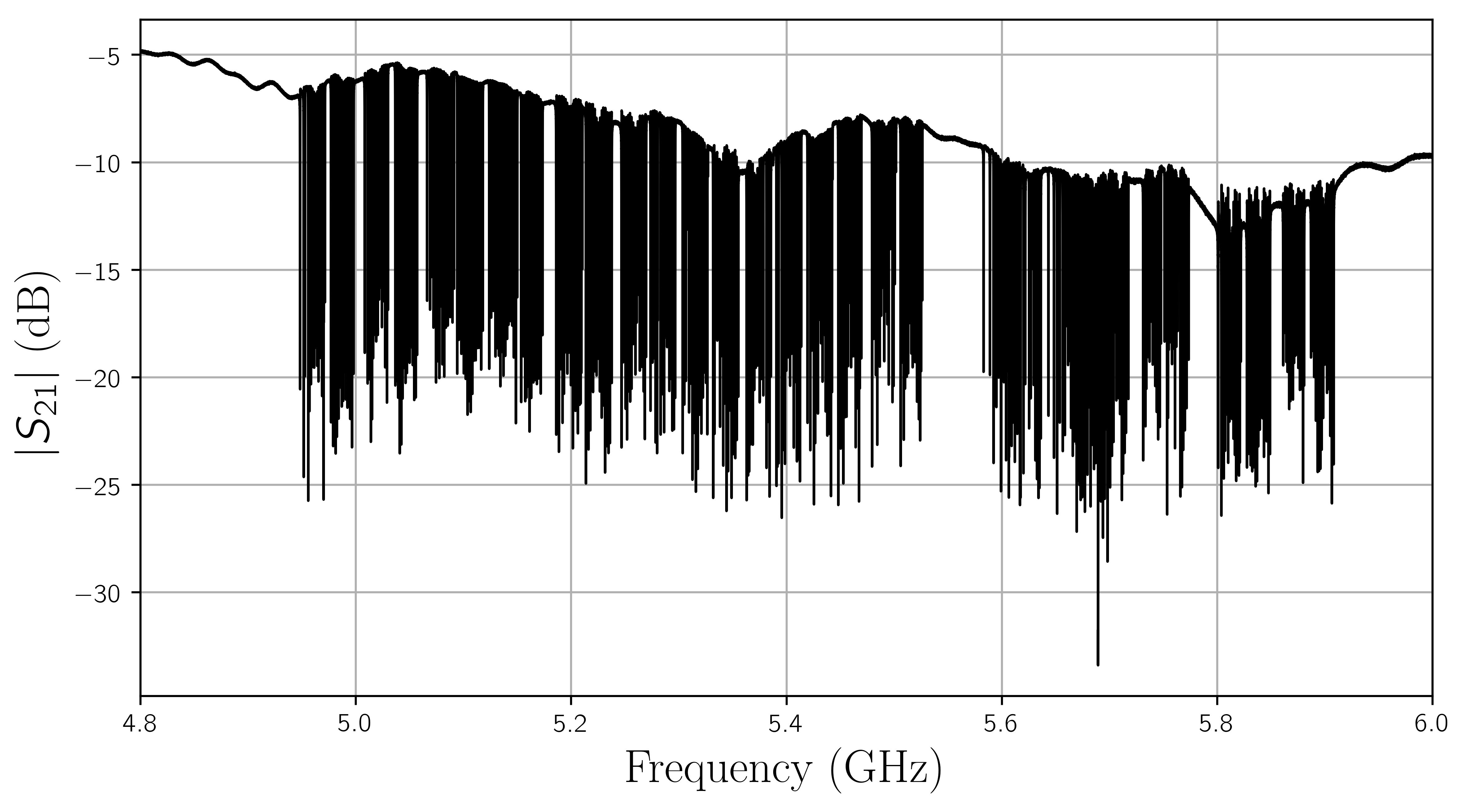}
	\caption{Measured transmission ($S_{21}$) through the NIST 528-channel microwave SQUID multiplexer at 250~mK from 4.8--6~GHz.  The resonator frequencies are intentionally grouped into four sub-bands per 66-channel multiplexer chip to guard against intra-chip resonator frequency collisions~\cite{DoberSPIE18}.  Two chips' resonator bands overlap in frequency between 5.576 and 5.778~GHz, likely due to wafer-scale variations in resonator properties which will be addressed in future designs.}
\end{figure}
The NIST 528-channel multiplexer consists of eight individual multiplexer chips (NIST mask ``umux100k"), each with 66 planar niobium coplanar waveguide (CPW) resonators capacitively coupled to a common CPW feedline~\cite{Bennett15}.  
To form the full 528-channel multiplexer the feedlines of the eight chips are connected in series, with each chip's resonators spanning a 125~MHz bandwidth.  
The 528 resonators of the full multiplexer package are distributed between 5--6~GHz with a measured median adjacent resonator frequency spacing of 1.5~MHz.  
This frequency spacing enables the placement of over 2000 resonator channels in the full planned SMuRF system bandwidth of 4~GHz.
Each individual multiplexer chip has one bare resonator channel and 65 rf-SQUID coupled resonators.
More details on the multiplexer including its design, characterization, and the RF packaging will be presented in a future paper~\cite{DoberSPIE18}.
Figure~\ref{fig:528vna} shows the measured transmission ($S_{21}$) through the 528-channel multiplexer using a vector network analyzer (VNA).

This VNA scan yields 512 easily identifiable resonators distributed between 4.94~GHz and 5.91~GHz. 
Two of the multiplexer chips were found to have a substantial overlap in their range of resonator frequencies, likely due to wafer-scale variations in resonator properties.
Of the 512 resonators identified from the VNA sweep, 477 of them fall into SMuRF's usable bandwidth between 5--6~GHz (those usable bands are $[5.002,5.498]$ GHz and $[5.502,5.998]$ GHz).
This multiplexer design did not consider the constraints imposed by the SMuRF electronics, but 
future designs will incorporate corrections for absolute frequency placement and wafer-scale variations in resonator properties.
These adjustments are needed to optimize the placement of the resonator frequencies for the SMuRF readout which is unable to track resonators which are too closely spaced or too near the edges of SMuRF's 500~MHz bands.

\subsection{Cryostat and cold RF circuit}
\label{subsec:cryostat}

To validate system performance, the multiplexer was mounted to the 250~mK stage  of a pulse tube cooled three-stage $^3$He sorption fridge. 
A single pair of coaxial cables carry signals between the multiplexer and the SMuRF, with thermal intercepts and additional RF components at the 350~mK, 2~K, 4~K, and 50~K stages of the cryostat.
9~dB of fixed cold attenuation in the cryostat helps to set the tone power levels into the $\mu$MUX resonators.
The input signal is coupled through the weakly coupled port of a 10~dB directional coupler~\footnote{Meca 780-10-6.000} mounted at the 350~mK stage in order to thermally isolate the multiplexer from the cold termination which dissipates the input signal power.
The power per tone which optimized the readout signal to noise was determined empirically to be approximately -75~dBm.

After transmission through the $\mu$MUX resonators, probe tones are coupled into the input of a two-stage amplifier chain through a cryogenic isolator at 350~mK.
The two-stage amplifier chain is designed to maximize the linearity of the cold RF circuit chain.
A low-gain high-electron-mobility transistor (HEMT)~\footnote{Low Noise Factory LNF-LNC4-8C\_LG} with noise temperature $T_N\sim 2K$ contributes +22~dB of gain at the 4~K stage.
A follow-on B\&Z Technologies cryogenic low noise amplifier (LNA)\footnote{BZUCR-04000800-031012-102323} at the 50~K stage contributes an additional +12.5~dB of gain.  
The physical temperature of the poorly named 50~K stage during normal operation was typically 38~K.

Figure~\ref{fig:linearity} shows the results of two-tone measurements of the linearity of this cryogenic two-stage amplifier chain versus the linearity of a single, high gain 4K HEMT with +39~dB gain~\footnote{Low Noise Factory LNF-LNC4-8C}.
The linearity for each amplifier chain is characterized by the third order intercept point (IP3), which is a measure of the input power level for which the amplified tones (Fundamental) are comparable in output power to the third order intermodulation distortion products (IMD3)~\cite{Pozar11}.  
The measured input IP3 (IIP3) for closely spaced tones near 5.5~GHz was -44.1~dBm for the single-stage cryogenic high gain HEMT and -29.5~dBm for the high linearity two-stage amplifier chain we have developed for use with SMuRF.
To prevent readout degradation due to cryogenic amplifier nonlinearity, the total power at the amplifier chain input must be well below the amplifier's IIP3.
For a typical attenuation on resonance of -15~dB and optimal tone power at the input to the resonators of -75~dBm, the summed amplifier input power for 2000 tones would be -57~dB, nearly 30~dB below the measured IIP3 of the high linearity two-stage cryogenic amplifier chain.

\subsection{SQUID tuning and tone tracking setup}
\label{subsec:tuning}
\begin{figure*}
\begin{center}
\resizebox{\textwidth}{!}{
\includegraphics[width=0.268\textwidth]{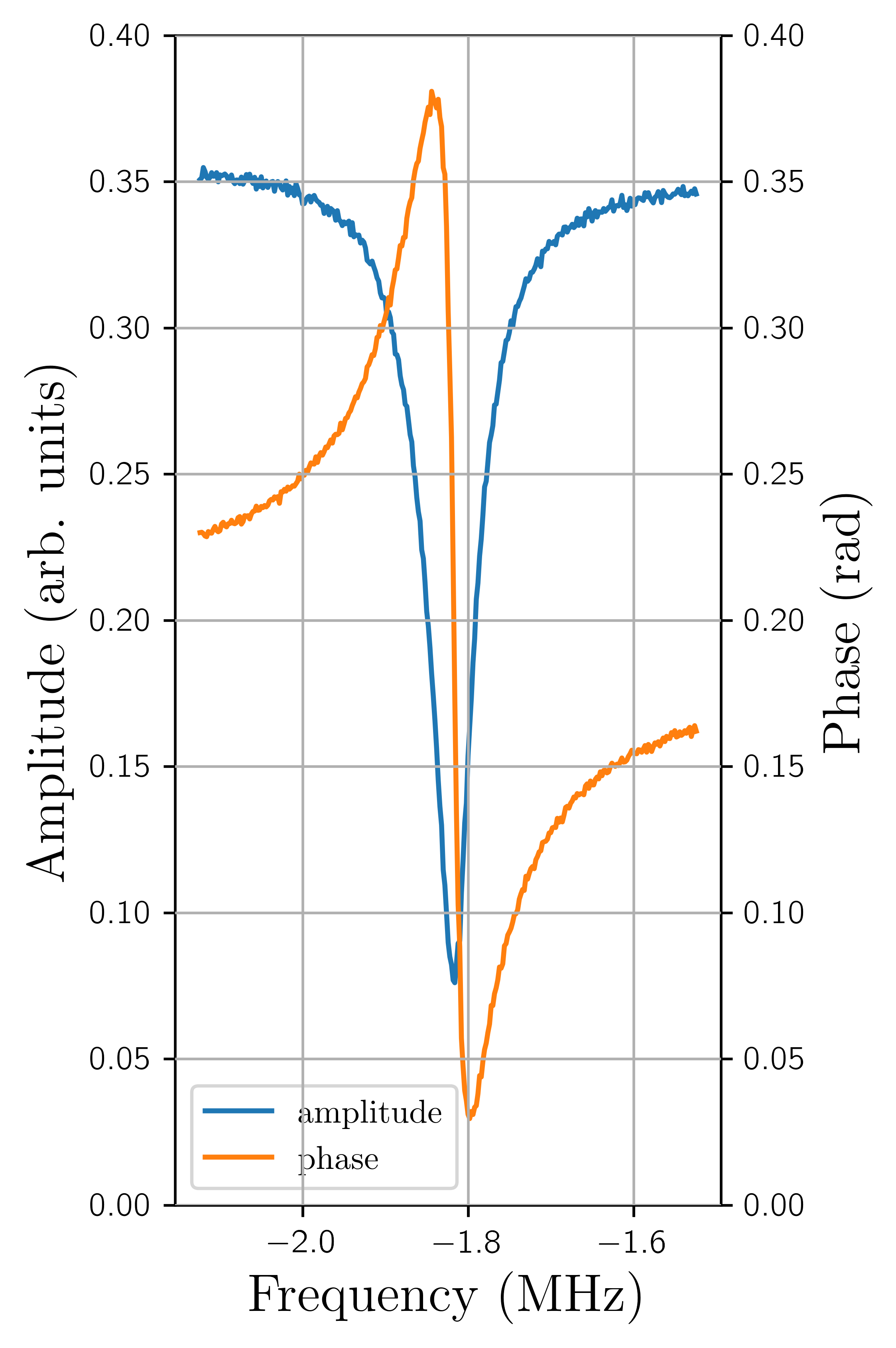}
\includegraphics[width=0.40\textwidth]{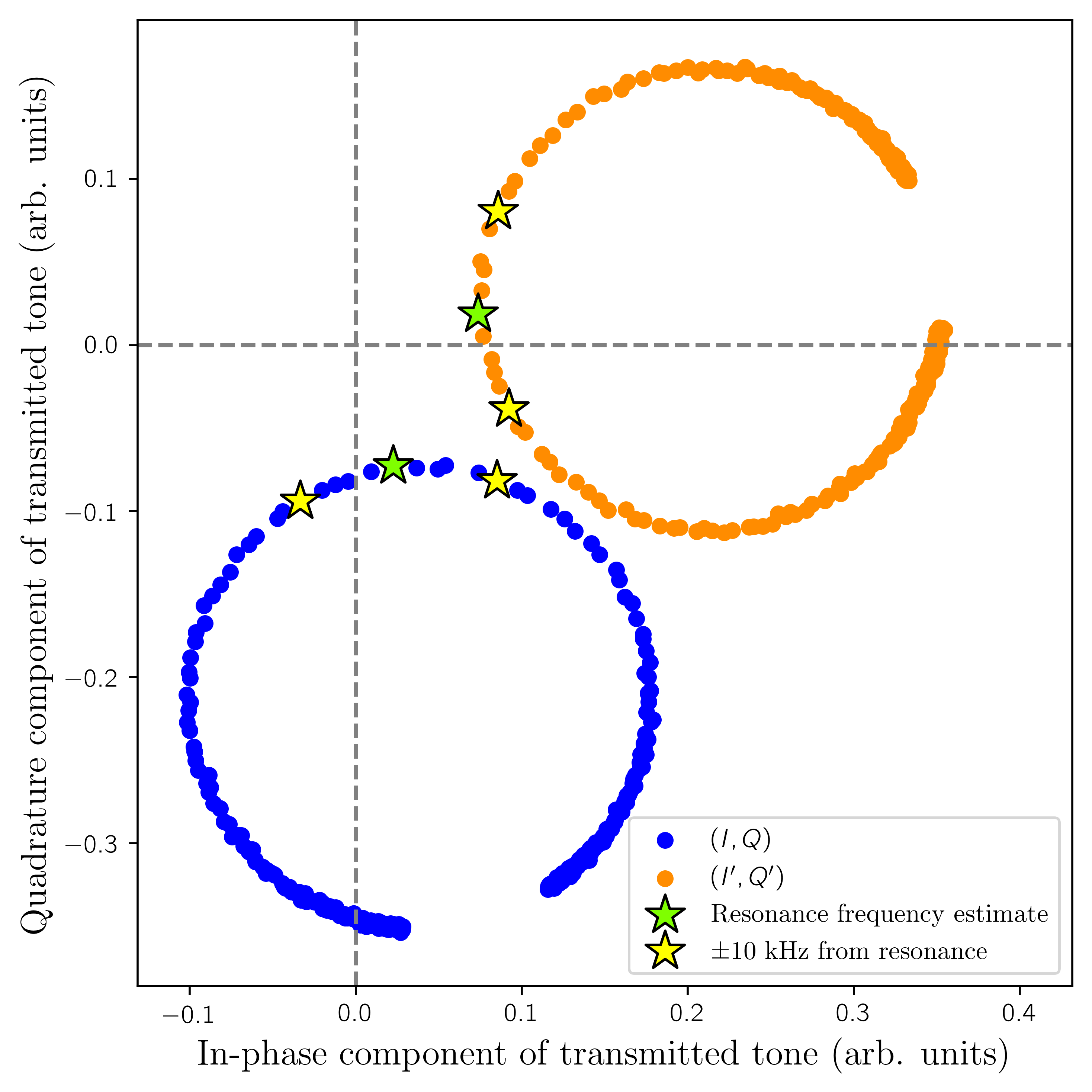}
\includegraphics[width=0.305\textwidth]{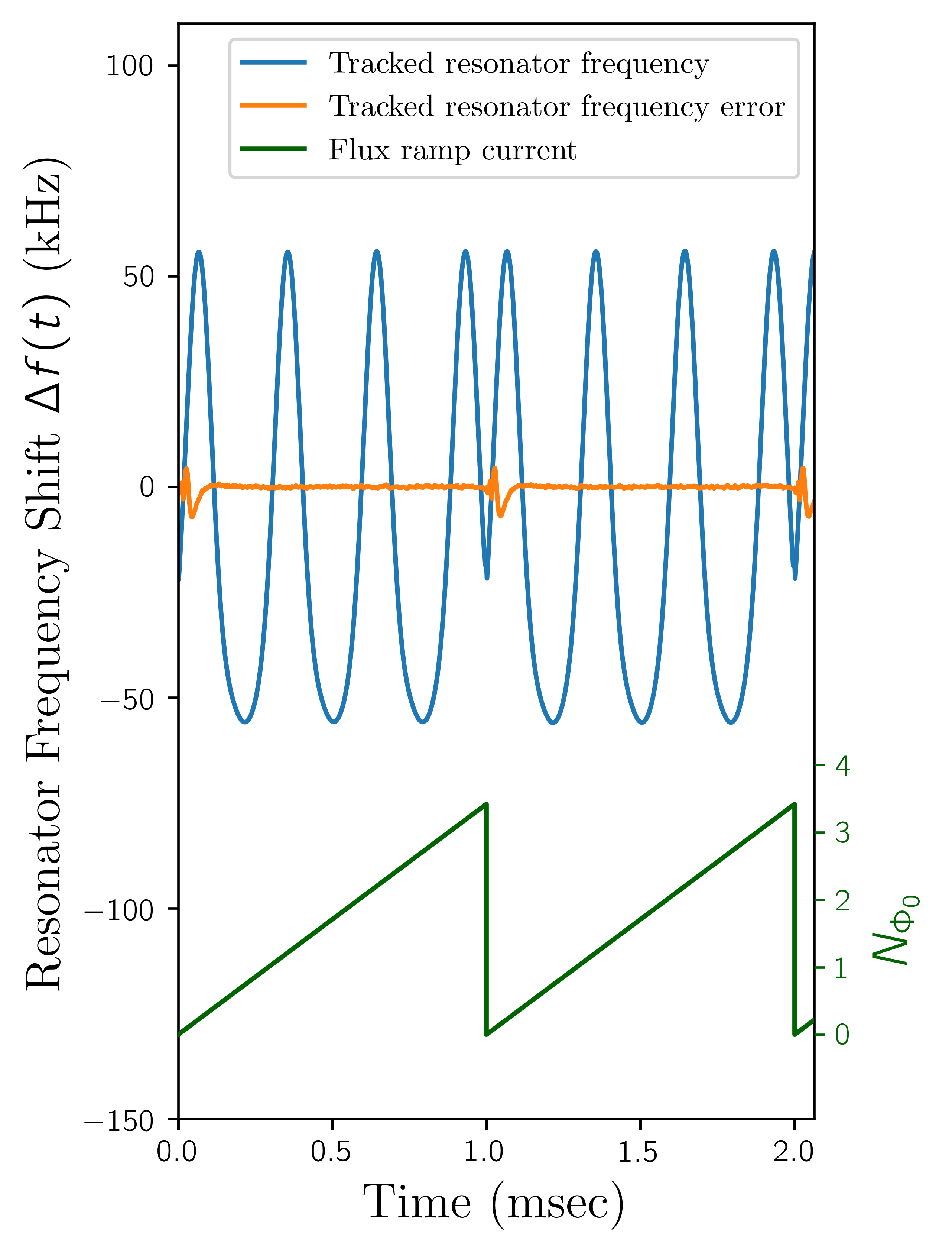}}
\end{center}
\caption{{\it Left:}  Transmitted probe tone amplitude (=$\sqrt{I^{2}+Q^{2}}$) and phase ($=\arctan{(Q/I)}$) versus probe tone frequency for a typical $\mu$MUX resonator in the 528-channel NIST multiplexer. 
{\it Center:}  The quadrature (Q) and in-phase (I) components of a transmitted probe tone as its frequency is swept over a typical $\mu$MUX resonance.  The transmitted tone traces a circle (blue) in the complex $IQ$ plane used to calibrate the SMuRF tracking algorithm as described in Section~\ref{subsec:tuning}.
Measurements of $(I,Q)$ on resonance (green stars) and at $\pm10$~kHz off resonance (yellow stars) are used to compute the tracked resonator frequency error using the scaled and rotated $I^{\prime}Q^{\prime}$ (orange) as described in Section~\ref{subsec:trackingalgo}.
{\it Right:} Tracked resonator frequency (blue) and tracked resonator frequency error (orange) versus time while applying a sawtooth flux ramp signal (green) injecting $\sim 3.4$ magnetic flux quanta $\Phi_{0}$ per period into the resonator's rf-SQUID at 1~kHz.  Within each flux ramp period, the resonator frequency follows the relationship given in Equation~\ref{eqn:freqshift}.}
\label{fig:tuning}
\end{figure*}
High-level control applications configure the SMuRF system to track and read out superconducting microwave resonators.  
We briefly outline here the procedures used to configure the SMuRF system, find each of the resonators' frequencies, set up closed loop frequency tracking for each resonance, and demodulate the flux modulation applied to the SQUIDs.

After power-on, the basic configuration defaults are first loaded, including which of each AMC RF cards' 500~MHz bands will be enabled.  
To identify the center frequency of each resonator, a single tone is then swept across the usable bandwidth of each 500~MHz SMuRF band in coarse frequency steps (typically 100~kHz), recording the in-phase and quadrature component of the transmitted tone $(I,Q)$ at each frequency point.  
A list of candidate resonator frequencies is determined from this raw response data by searching for characteristic jumps in the phase ($=\arctan(Q/I)$) of the transmitted tone as a function of tone frequency.

A single tone is then swept more finely about each candidate resonance frequency (typically in 2~kHz steps over a frequency range of $\pm300$~kHz) in order to map the transfer function near a resonance~\cite{Zmuidzinas12}.
The transmitted tone amplitude ($=\sqrt{I^{2}+Q^{2}}$), phase, and complex response $(I,Q)$ versus tone frequency for a typical resonator is shown in Figure~\ref{fig:tuning}.
Each resonator's resonance frequency $f_{r}$ is estimated by finding the frequency for which the transmitted tone amplitude is smallest.
The transformation from complex coordinates $(I,Q)$ to $(I^{\prime},Q^{\prime})$ required as input to the tracking algorithm (see Section~\ref{subsec:trackingalgo}) is estimated as the inverse of the unit modulus difference in complex transmission measured at a frequency offset $\Delta f_{\pm}$ on either side of the resonance frequency $(I_{\pm},Q_{\pm})$.  
$\Delta f_{\pm}$ was empirically chosen to be $10$~kHz, as shown in Figure~\ref{fig:tuning}.
For each resonance the SMuRF tracking algorithm takes as input the complex parameter $\overline{\eta}=2\Delta f_{\pm}( (I_{+}-I_{-})+i (Q_{+}-Q_{-}))^{-1}$ derived from these measurements.
Near resonance, $\operatorname{Im}[\overline{\eta}\times(I,Q)]\sim\Delta f$, the tracked frequency error.

Within each 500~MHz SMuRF band a tone is assigned to each resonance by programming firmware tone blocks with each resonator's $f_{r}$ and $\overline{\eta}$ parameter. 
Once the individual tones are programmed and assigned, the RTM is commanded to generate a sawtooth flux ramp signal with a specific frequency and amplitude.
SMuRF's tracking algorithm is configured by specifying the carrier frequency $f_{c}$ common to all channels and a single gain term (for more details see Section~\ref{subsec:trackingalgo}).
The carrier frequency $f_{c}$ is determined empirically by minimizing the tracked frequency error.

Once tones are assigned and tracking the flux ramp modulated resonators, for each resonator the system reports a short snapshot of tracked resonator frequency and resonator frequency error versus time as shown in Figure~\ref{fig:tuning}.  
Channels are disabled if they are not tracking properly, or have atypical tracked frequency versus time data.
After this quality check the system is instructed to report the instantaneous demodulated phase of the first harmonic for each tracking channel, which is proportional to the flux in each resonator's rf-SQUID.

\subsection{Noise performance}
\label{subsec:noise}
After following the procedure outlined in Section~\ref{subsec:tuning}, SMuRF is able to track 426 of the 477 resonators whose frequencies fall in SMuRF's usable 5--6~GHz bandwidth.  
7 of the channels that the system is unable to track are the dark resonator channels which fall in SMuRF's usable bandwidth which are intentionally not coupled to rf-SQUIDs and subsequently not modulated by the flux ramp.  
Of the remaining 44 resonators which SMuRF is unable to track, the majority ($\sim$36) are adjacent resonators which are near or closer to one another than a resonator line width, ($\sim100$~kHz).
The majority of these colliding resonator pairs fall in the range of frequencies over which two multiplexer chips' resonators unintentionally overlap in frequency, as described in Section~\ref{subsec:NIST528}.  
Finally, $\sim$8 channels are disabled in the quality cuts described in Section~\ref{subsec:tuning} typically either due to the resonance having an anomalous measured IQ circle (often due to a low quality factor) or an atypical response to flux ramp modulation. 

\begin{figure}[!hbt]
	\centering
	\label{fig:528noise}
	\includegraphics[width=\textwidth]{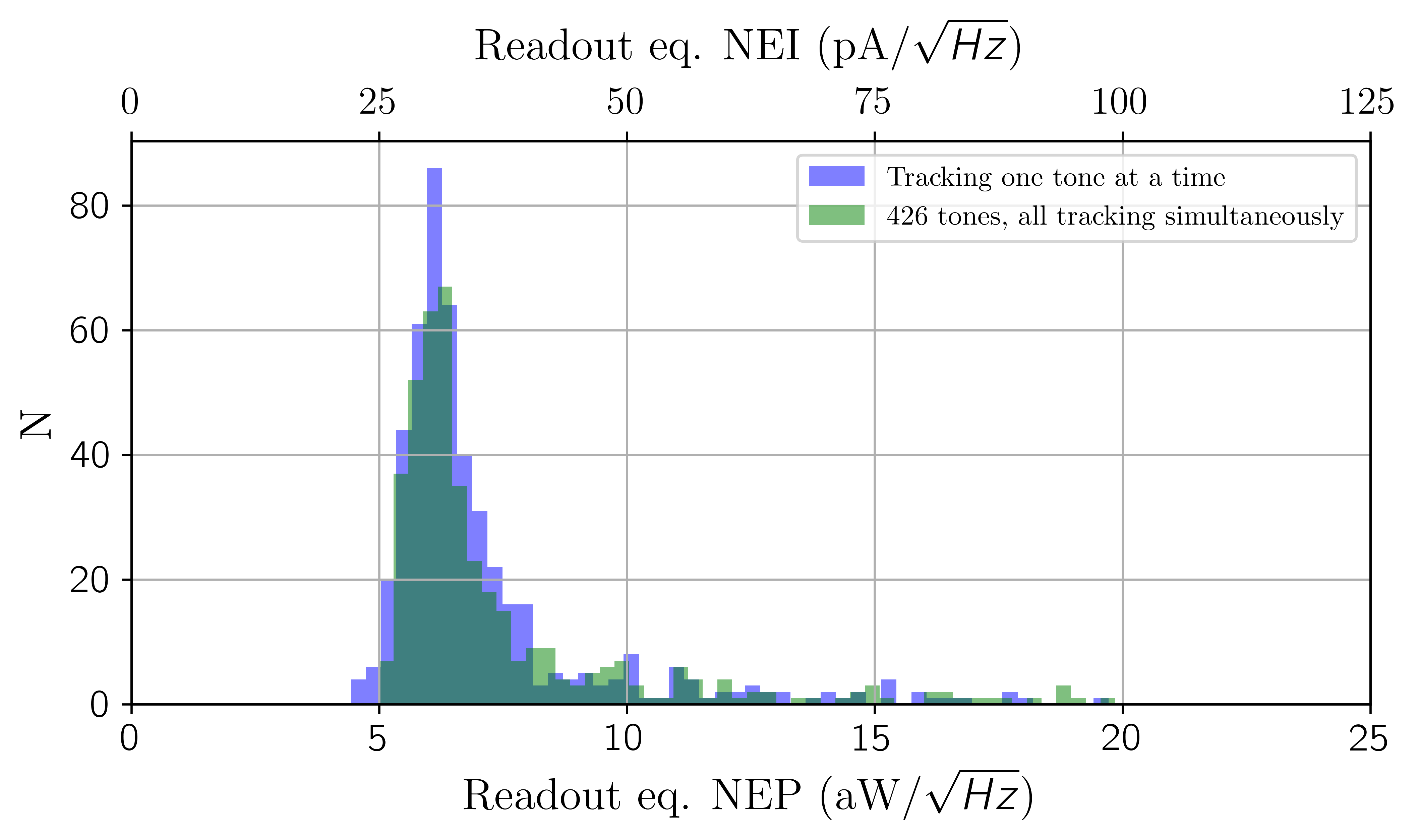}
	\caption{Measured TES-equivalent readout white noise level in noise equivalent current (NEI) and noise equivalent power (NEP) for SMuRF tracking 426 resonators in the NIST 528 channel $\mu$MUX multiplexer.  The white noise level for each resonance is estimated as the median of the current noise amplitude spectral density of a 60~second dataset in the $[1,10]$~Hz signal band.  Noise taken with only one channel on and tracking at a time for each resonator agrees with noise taken with all 426 channels tracking simultaneously.}
\end{figure}

Figure~\ref{fig:528noise} is a histogram of the measured readout white noise level for all 426 tracking channels in both noise equivalent TES current (NEI) and power (NEP). 
For these measurements, the flux ramp reset rate was chosen to be 4~kHz with $\sim$3.3 magnetic flux quanta swept in each flux ramp period, resulting in a $\sim$13~kHz carrier rate.
No TESs were connected to the inputs of the multiplexer for this measurement, so although we report equivalent TES noise the noise is sourced by the readout alone.  
The noise was measured in two configurations: 1) with all channels tracking simultaneously, and 2) with only one channel tracking at a time.  
Each noise measurement is derived from a 60 second dataset, sampling each channel at 600~kHz.  
The SMuRF tracking algorithm reports the demodulated phase $\phi_{demod}$ in radians for each channel at this data rate, which is related to equivalent TES current through

\begin{equation}
I_{TES}=\frac{\phi_{demod}\Phi_{0}}{2\pi M_{in}}
\label{eq:rad2pA}
\end{equation}

where $I_{TES}$ is the equivalent TES current, $\Phi_{0}$ is the magnetic flux quantum, and $M_{in}$ is the mutual inductance between the TES loop and its rf-SQUID~\cite{Mates11PhD}.  
The design value of $M_{in}$ for this multiplexer is 228~pH which we verified to be correct to within 5\% by injecting a known current directly into the input coil of one of the rf-SQUIDs in the multiplexer and using SMuRF to measure the equivalent phase in radians.

For each resonator we compute the amplitude spectral density of its demodulated phase versus time data in rad$/\sqrt{\mathrm{Hz}}$.  We convert this phase amplitude spectral density into an equivalent TES current amplitude spectral density in pA$/\sqrt{\mathrm{Hz}}$ using Equation~\ref{eq:rad2pA}.  We then estimate the equivalent white TES current noise level (NEI) for each channel by taking the median over the $[1,10]$~Hz frequency interval of the current amplitude spectral density.  The equivalent readout NEP in Figure~\ref{fig:528noise} is then computed from the NEI estimate for each resonator from

\begin{equation}
\mathrm{NEP} = \sqrt{P_{bias}R_{TES}}\times\mathrm{NEI}
\label{eq:NEI2NEP}
\end{equation}

where $P_{bias}$ is the electrical power dissipated in the TES due to its voltage bias and $R_{TES}$ is the operating resistance of the TES.  As an example, we assume $R_{TES}=4$ m$\Omega$ and $P_{bias}=10$ pW, values typical of TESs observing the CMB from the Atacama at 90 and 150~GHz~\cite{Choi18}.

The median NEI (NEP) over all 426 channels from Figure~\ref{fig:528noise} is $32$ pA$/\sqrt{\mathrm{Hz}}$ ($6.4$ aW$/\sqrt{\mathrm{Hz}}$) and is consistent within 1\% between data taken with only one channel tracking at a time and data taken with all 426 channels tracking simultaneously.
In addition to being the highest achieved $\mu$MUX multiplexing factor to date this measurement confirms, as expected, that the added readout noise due to nonlinearities in the cold RF electronics and the SMuRF system when reading out 426 channels contributes negligibly to the total readout noise.
The total measured readout noise also agrees with independent measurements of the noise on single resonators on chips from the same wafer used to fabricate the 528 channel multiplexer used for this demonstration using the NIST ROACH-2 based $\mu$MUX readout electronics~\cite{DoberSPIE18}.

\section{Conclusion}

Microwave SQUID multiplexing ($\mu$MUX) has been identified as a potentially enabling technology for CMB-S4~\cite{CMBs4Science}.
In particular, the CMB-S4 Technology Book specifically identifies the need for the development of a universal MKID and $\mu$MUX warm readout with resonator tone tracking~\cite{CMBs4Tech}.
We are developing the SLAC Microwave RF Electronics (SMuRF) to meet these needs for the next generation of CMB and sub-mm surveys.  
Using a mature SMuRF prototype operating on 1~GHz of bandwidth and a cold RF chain optimized for linearity we have demonstrated the highest $\mu$MUX multiplexing factor yet achieved of 426 channels, reading out a 528-channel $\mu$MUX cryogenic multiplexer fabricated at NIST Boulder~\cite{DoberSPIE18}.

The measured readout white noise level is consistent with expectations and well suited to the readout noise requirements of existing and future ground-based mm- and sub-mm observatories.
Further exploration of the readout system performance including low-frequency noise, crosstalk, linearity, and operation with TESs is underway and will be detailed in future publications.
Based on the measured noise levels with 426 tones tracking and the achieved linearity of the cold RF chain and SMuRF electronics detailed in these proceedings, we expect to be able to achieve multiplexing factors exceeding 2000 with the full SMuRF system and readout noise levels below the requirements of future planned CMB and sub-mm efforts.
The full SMuRF system will operate over a 4~GHz bandwidth from 4--8~GHz and 
will be capable of tracking more than 4000 tones.  
As will be described in a future paper~\cite{DoberSPIE18} this work is well complemented by the on-going development of a 2000-channel cryogenic $\mu$MUX multiplexer at NIST Boulder with resonators distributed in the SMuRF 4--8~GHz band which builds on the design of the 528-channel cryogenic multiplexer tested in these proceedings.

A new revision of the SMuRF hardware in production will upgrade the FPGA on the carrier card to a KU15P Ultrascale+, enabling tone tracking on 4000 $\mu$MUX channels over the full 4--8~GHz bandwidth.
In addition to CMB-S4, SMuRF is being developed as a possible readout solution for the Simons Observatory~\cite{Galitzki18}, CCAT-prime~\cite{CCATprimeSPIE18}, BICEP Array~\cite{Grayson16}, and Ali-CPT~\cite{AliCPT18} projects.  SMuRF is simultaneously being developed to read out X-ray microcalorimeter arrays for the LCLS-II experiment~\cite{Frisch17,Li18}.

\acknowledgments 
This work was supported by the U.S. Department of Energy, Office of Science, Contract DE-AC02-76SF00515 and FWP 2017-SLAC-100334.
CY is supported by a National Science Foundation Graduate Research Fellowship.

\bibliography{report}

\begin{thebibliography}{10}

\bibitem{CMBs4Science}
K.~N. {Abazajian}, P.~{Adshead}, Z.~{Ahmed}, S.~W. {Allen}, D.~{Alonso}, K.~S.
  {Arnold}, C.~{Baccigalupi}, J.~G. {Bartlett}, N.~{Battaglia}, B.~A. {Benson},
  C.~A. {Bischoff}, J.~{Borrill}, V.~{Buza}, E.~{Calabrese}, R.~{Caldwell},
  J.~E. {Carlstrom}, C.~L. {Chang}, T.~M. {Crawford}, F.-Y. {Cyr-Racine},
  F.~{De Bernardis}, T.~{de Haan}, S.~{di Serego Alighieri}, J.~{Dunkley},
  C.~{Dvorkin}, J.~{Errard}, G.~{Fabbian}, S.~{Feeney}, S.~{Ferraro}, J.~P.
  {Filippini}, R.~{Flauger}, G.~M. {Fuller}, V.~{Gluscevic}, D.~{Green},
  D.~{Grin}, E.~{Grohs}, J.~W. {Henning}, J.~C. {Hill}, R.~{Hlozek},
  G.~{Holder}, W.~{Holzapfel}, W.~{Hu}, K.~M. {Huffenberger}, R.~{Keskitalo},
  L.~{Knox}, A.~{Kosowsky}, J.~{Kovac}, E.~D. {Kovetz}, C.-L. {Kuo},
  A.~{Kusaka}, M.~{Le Jeune}, A.~T. {Lee}, M.~{Lilley}, M.~{Loverde}, M.~S.
  {Madhavacheril}, A.~{Mantz}, D.~J.~E. {Marsh}, J.~{McMahon}, P.~D.
  {Meerburg}, J.~{Meyers}, A.~D. {Miller}, J.~B. {Munoz}, H.~N. {Nguyen}, M.~D.
  {Niemack}, M.~{Peloso}, J.~{Peloton}, L.~{Pogosian}, C.~{Pryke}, M.~{Raveri},
  C.~L. {Reichardt}, G.~{Rocha}, A.~{Rotti}, E.~{Schaan}, M.~M. {Schmittfull},
  D.~{Scott}, N.~{Sehgal}, S.~{Shandera}, B.~D. {Sherwin}, T.~L. {Smith},
  L.~{Sorbo}, G.~D. {Starkman}, K.~T. {Story}, A.~{van Engelen}, J.~D.
  {Vieira}, S.~{Watson}, N.~{Whitehorn}, and W.~L. {Kimmy Wu}, ``{CMB-S4
  Science Book, First Edition},'' {\em ArXiv e-prints} , Oct. 2016, 1610.02743.

\bibitem{CMBs4Tech}
M.~H. {Abitbol}, Z.~{Ahmed}, D.~{Barron}, R.~{Basu Thakur}, A.~N. {Bender},
  B.~A. {Benson}, C.~A. {Bischoff}, S.~A. {Bryan}, J.~E. {Carlstrom}, C.~L.
  {Chang}, D.~T. {Chuss}, K.~T. {Crowley}, A.~{Cukierman}, T.~{de Haan},
  M.~{Dobbs}, T.~{Essinger-Hileman}, J.~P. {Filippini}, K.~{Ganga}, J.~E.
  {Gudmundsson}, N.~W. {Halverson}, S.~{Hanany}, S.~W. {Henderson}, C.~A.
  {Hill}, S.-P.~P. {Ho}, J.~{Hubmayr}, K.~{Irwin}, O.~{Jeong}, B.~R. {Johnson},
  S.~A. {Kernasovskiy}, J.~M. {Kovac}, A.~{Kusaka}, A.~T. {Lee}, S.~{Maria},
  P.~{Mauskopf}, J.~J. {McMahon}, L.~{Moncelsi}, A.~W. {Nadolski}, J.~M.
  {Nagy}, M.~D. {Niemack}, R.~C. {O'Brient}, S.~{Padin}, S.~C. {Parshley},
  C.~{Pryke}, N.~A. {Roe}, K.~{Rostem}, J.~{Ruhl}, S.~M. {Simon}, S.~T.
  {Staggs}, A.~{Suzuki}, E.~R. {Switzer}, O.~{Tajima}, K.~L. {Thompson},
  P.~{Timbie}, G.~S. {Tucker}, J.~D. {Vieira}, A.~G. {Vieregg}, B.~{Westbrook},
  E.~J. {Wollack}, K.~W. {Yoon}, K.~S. {Young}, and E.~Y. {Young}, ``{CMB-S4
  Technology Book, First Edition},'' {\em ArXiv e-prints} , June 2017,
  1706.02464.

\bibitem{AdvACT16}
S.~W. {Henderson}, R.~{Allison}, J.~{Austermann}, T.~{Baildon}, N.~{Battaglia},
  J.~A. {Beall}, D.~{Becker}, F.~{De Bernardis}, J.~R. {Bond}, E.~{Calabrese},
  S.~K. {Choi}, K.~P. {Coughlin}, K.~T. {Crowley}, R.~{Datta}, M.~J. {Devlin},
  S.~M. {Duff}, J.~{Dunkley}, R.~{D{\"u}nner}, A.~{van Engelen}, P.~A.
  {Gallardo}, E.~{Grace}, M.~{Hasselfield}, F.~{Hills}, G.~C. {Hilton}, A.~D.
  {Hincks}, R.~{Hlo{\^z}ek}, S.~P. {Ho}, J.~{Hubmayr}, K.~{Huffenberger}, J.~P.
  {Hughes}, K.~D. {Irwin}, B.~J. {Koopman}, A.~B. {Kosowsky}, D.~{Li},
  J.~{McMahon}, C.~{Munson}, F.~{Nati}, L.~{Newburgh}, M.~D. {Niemack},
  P.~{Niraula}, L.~A. {Page}, C.~G. {Pappas}, M.~{Salatino}, A.~{Schillaci},
  B.~L. {Schmitt}, N.~{Sehgal}, B.~D. {Sherwin}, J.~L. {Sievers}, S.~M.
  {Simon}, D.~N. {Spergel}, S.~T. {Staggs}, J.~R. {Stevens}, R.~{Thornton},
  J.~{Van Lanen}, E.~M. {Vavagiakis}, J.~T. {Ward}, and E.~J. {Wollack},
  ``{Advanced ACTPol Cryogenic Detector Arrays and Readout},'' {\em Journal of
  Low Temperature Physics}~{\bf 184}, pp.~772--779, Aug. 2016, 1510.02809.

\bibitem{Grayson16}
J.~A. {Grayson}, P.~A.~R. {Ade}, Z.~{Ahmed}, K.~D. {Alexander}, M.~{Amiri},
  D.~{Barkats}, S.~J. {Benton}, C.~A. {Bischoff}, J.~J. {Bock}, H.~{Boenish},
  R.~{Bowens-Rubin}, I.~{Buder}, E.~{Bullock}, V.~{Buza}, J.~{Connors}, J.~P.
  {Filippini}, S.~{Fliescher}, M.~{Halpern}, S.~{Harrison}, G.~C. {Hilton},
  V.~V. {Hristov}, H.~{Hui}, K.~D. {Irwin}, J.~{Kang}, K.~S. {Karkare},
  E.~{Karpel}, S.~{Kefeli}, S.~A. {Kernasovskiy}, J.~M. {Kovac}, C.~L. {Kuo},
  E.~M. {Leitch}, M.~{Lueker}, K.~G. {Megerian}, V.~{Monticue}, T.~{Namikawa},
  C.~B. {Netterfield}, H.~T. {Nguyen}, R.~{O'Brient}, R.~W. {Ogburn},
  C.~{Pryke}, C.~D. {Reintsema}, S.~{Richter}, R.~{Schwarz}, C.~{Sorenson},
  C.~D. {Sheehy}, Z.~K. {Staniszewski}, B.~{Steinbach}, G.~P. {Teply}, K.~L.
  {Thompson}, J.~E. {Tolan}, C.~{Tucker}, A.~D. {Turner}, A.~G. {Vieregg},
  A.~{Wandui}, A.~C. {Weber}, D.~V. {Wiebe}, J.~{Willmert}, W.~L.~K. {Wu}, and
  K.~W. {Yoon}, ``{BICEP3 performance overview and planned Keck Array
  upgrade},'' in {\em Millimeter, Submillimeter, and Far-Infrared Detectors and
  Instrumentation for Astronomy VIII},  {\em Proc. SPIE} {\bf 9914}, p.~99140S,
  July 2016, 1607.04668.

\bibitem{Suzuki16}
A.~{Suzuki}, P.~{Ade}, Y.~{Akiba}, C.~{Aleman}, K.~{Arnold}, C.~{Baccigalupi},
  B.~{Barch}, D.~{Barron}, A.~{Bender}, D.~{Boettger}, J.~{Borrill},
  S.~{Chapman}, Y.~{Chinone}, A.~{Cukierman}, M.~{Dobbs}, A.~{Ducout},
  R.~{Dunner}, T.~{Elleflot}, J.~{Errard}, G.~{Fabbian}, S.~{Feeney},
  C.~{Feng}, T.~{Fujino}, G.~{Fuller}, A.~{Gilbert}, N.~{Goeckner-Wald},
  J.~{Groh}, T.~D. {Haan}, G.~{Hall}, N.~{Halverson}, T.~{Hamada},
  M.~{Hasegawa}, K.~{Hattori}, M.~{Hazumi}, C.~{Hill}, W.~{Holzapfel},
  Y.~{Hori}, L.~{Howe}, Y.~{Inoue}, F.~{Irie}, G.~{Jaehnig}, A.~{Jaffe},
  O.~{Jeong}, N.~{Katayama}, J.~{Kaufman}, K.~{Kazemzadeh}, B.~{Keating},
  Z.~{Kermish}, R.~{Keskitalo}, T.~{Kisner}, A.~{Kusaka}, M.~L. {Jeune},
  A.~{Lee}, D.~{Leon}, E.~{Linder}, L.~{Lowry}, F.~{Matsuda}, T.~{Matsumura},
  N.~{Miller}, K.~{Mizukami}, J.~{Montgomery}, M.~{Navaroli}, H.~{Nishino},
  J.~{Peloton}, D.~{Poletti}, G.~{Puglisi}, G.~{Rebeiz}, C.~{Raum},
  C.~{Reichardt}, P.~{Richards}, C.~{Ross}, K.~{Rotermund}, Y.~{Segawa},
  B.~{Sherwin}, I.~{Shirley}, P.~{Siritanasak}, N.~{Stebor}, R.~{Stompor},
  J.~{Suzuki}, O.~{Tajima}, S.~{Takada}, S.~{Takakura}, S.~{Takatori},
  A.~{Tikhomirov}, T.~{Tomaru}, B.~{Westbrook}, N.~{Whitehorn}, T.~{Yamashita},
  A.~{Zahn}, and O.~{Zahn}, ``{The Polarbear-2 and the Simons Array
  Experiments},'' {\em Journal of Low Temperature Physics}~{\bf 184},
  pp.~805--810, Aug. 2016, 1512.07299.

\bibitem{Anderson2018}
A.~J. Anderson, P.~A.~R. Ade, Z.~Ahmed, J.~E. Austermann, J.~S. Avva, P.~S.
  Barry, R.~B. Thakur, A.~N. Bender, B.~A. Benson, L.~E. Bleem, K.~Byrum, J.~E.
  Carlstrom, F.~W. Carter, T.~Cecil, C.~L. Chang, H.~M. Cho, J.~F. Cliche,
  T.~M. Crawford, A.~Cukierman, E.~V. Denison, T.~de~Haan, J.~Ding, M.~A.
  Dobbs, D.~Dutcher, W.~Everett, A.~Foster, R.~N. Gannon, A.~Gilbert, J.~C.
  Groh, N.~W. Halverson, A.~H. Harke-Hosemann, N.~L. Harrington, J.~W. Henning,
  G.~C. Hilton, G.~P. Holder, W.~L. Holzapfel, N.~Huang, K.~D. Irwin, O.~B.
  Jeong, M.~Jonas, T.~Khaire, L.~Knox, A.~M. Kofman, M.~Korman, D.~Kubik,
  S.~Kuhlmann, N.~Kuklev, C.~L. Kuo, A.~T. Lee, E.~M. Leitch, A.~E. Lowitz,
  S.~S. Meyer, D.~Michalik, J.~Montgomery, A.~Nadolski, T.~Natoli, H.~Nguyen,
  G.~I. Noble, V.~Novosad, S.~Padin, Z.~Pan, J.~Pearson, C.~M. Posada,
  A.~Rahlin, C.~L. Reichardt, J.~E. Ruhl, L.~J. Saunders, J.~T. Sayre,
  I.~Shirley, E.~Shirokoff, G.~Smecher, J.~A. Sobrin, A.~A. Stark, K.~T. Story,
  A.~Suzuki, Q.~Y. Tang, K.~L. Thompson, C.~Tucker, L.~R. Vale, K.~Vanderlinde,
  J.~D. Vieira, G.~Wang, N.~Whitehorn, V.~Yefremenko, K.~W. Yoon, and M.~R.
  Young, ``{SPT-3G: A Multichroic Receiver for the South Pole Telescope},''
  {\em Journal of Low Temperature Physics} , Jul 2018.

\bibitem{Duff16}
S.~M. {Duff}, J.~{Austermann}, J.~A. {Beall}, D.~{Becker}, R.~{Datta}, P.~A.
  {Gallardo}, S.~W. {Henderson}, G.~C. {Hilton}, S.~P. {Ho}, J.~{Hubmayr},
  B.~J. {Koopman}, D.~{Li}, J.~{McMahon}, F.~{Nati}, M.~D. {Niemack}, C.~G.
  {Pappas}, M.~{Salatino}, B.~L. {Schmitt}, S.~M. {Simon}, S.~T. {Staggs},
  J.~R. {Stevens}, J.~{Van Lanen}, E.~M. {Vavagiakis}, J.~T. {Ward}, and E.~J.
  {Wollack}, ``{Advanced ACTPol Multichroic Polarimeter Array Fabrication
  Process for 150 mm Wafers},'' {\em Journal of Low Temperature Physics}~{\bf
  184}, pp.~634--641, Aug. 2016.

\bibitem{Posada15}
C.~M. {Posada}, P.~A.~R. {Ade}, Z.~{Ahmed}, K.~{Arnold}, J.~E. {Austermann},
  A.~N. {Bender}, L.~E. {Bleem}, B.~A. {Benson}, K.~{Byrum}, J.~E. {Carlstrom},
  C.~L. {Chang}, H.~M. {Cho}, S.~T. {Ciocys}, J.~F. {Cliche}, T.~M. {Crawford},
  A.~{Cukierman}, D.~{Czaplewski}, J.~{Ding}, R.~{Divan}, T.~{de Haan}, M.~A.
  {Dobbs}, D.~{Dutcher}, W.~{Everett}, A.~{Gilbert}, N.~W. {Halverson}, N.~L.
  {Harrington}, K.~{Hattori}, J.~W. {Henning}, G.~C. {Hilton}, W.~L.
  {Holzapfel}, J.~{Hubmayr}, K.~D. {Irwin}, O.~{Jeong}, R.~{Keisler},
  D.~{Kubik}, C.~L. {Kuo}, A.~T. {Lee}, E.~M. {Leitch}, S.~{Lendinez}, S.~S.
  {Meyer}, C.~S. {Miller}, J.~{Montgomery}, M.~{Myers}, A.~{Nadolski},
  T.~{Natoli}, H.~{Nguyen}, V.~{Novosad}, S.~{Padin}, Z.~{Pan}, J.~{Pearson},
  J.~E. {Ruhl}, B.~R. {Saliwanchik}, G.~{Smecher}, J.~T. {Sayre},
  E.~{Shirokoff}, L.~{Stan}, A.~A. {Stark}, J.~{Sobrin}, K.~{Story},
  A.~{Suzuki}, K.~L. {Thompson}, C.~{Tucker}, K.~{Vanderlinde}, J.~D. {Vieira},
  G.~{Wang}, N.~{Whitehorn}, V.~{Yefremenko}, K.~W. {Yoon}, and K.~E.
  {Ziegler}, ``{Fabrication of large dual-polarized multichroic TES bolometer
  arrays for CMB measurements with the SPT-3G camera},'' {\em Superconductor
  Science Technology}~{\bf 28}, p.~094002, Sept. 2015.

\bibitem{Suzuki12}
A.~{Suzuki}, K.~{Arnold}, J.~{Edwards}, G.~{Engargiola}, A.~{Ghribi},
  W.~{Holzapfel}, A.~{Lee}, X.~{Meng}, M.~{Myers}, R.~{O'Brient}, E.~{Quealy},
  G.~{Rebeiz}, and P.~{Richards}, ``{Multi-chroic Dual-Polarization Bolometric
  Focal Plane for Studies of the Cosmic Microwave Background},'' {\em Journal
  of Low Temperature Physics} , pp.~852--858, June 2012, 1210.8256.

\bibitem{Henderson16}
S.~W. {Henderson}, J.~R. {Stevens}, M.~{Amiri}, J.~{Austermann}, J.~A. {Beall},
  S.~{Chaudhuri}, H.-M. {Cho}, S.~K. {Choi}, N.~F. {Cothard}, K.~T. {Crowley},
  S.~M. {Duff}, C.~P. {Fitzgerald}, P.~A. {Gallardo}, M.~{Halpern},
  M.~{Hasselfield}, G.~{Hilton}, S.-P.~P. {Ho}, J.~{Hubmayr}, K.~D. {Irwin},
  B.~J. {Koopman}, D.~{Li}, Y.~{Li}, J.~{McMahon}, F.~{Nati}, M.~{Niemack},
  C.~D. {Reintsema}, M.~{Salatino}, A.~{Schillaci}, B.~L. {Schmitt}, S.~M.
  {Simon}, S.~T. {Staggs}, E.~M. {Vavagiakis}, and J.~T. {Ward}, ``{Readout of
  two-kilopixel transition-edge sensor arrays for Advanced ACTPol},'' in {\em
  Millimeter, Submillimeter, and Far-Infrared Detectors and Instrumentation for
  Astronomy VIII},  {\em Proc. SPIE} {\bf 9914}, p.~99141G, July 2016,
  1607.06064.

\bibitem{Bender14}
A.~N. {Bender}, J.-F. {Cliche}, T.~{de Haan}, M.~A. {Dobbs}, A.~J. {Gilbert},
  J.~{Montgomery}, N.~{Rowlands}, G.~M. {Smecher}, K.~{Smith}, and A.~{Wilson},
  ``{Digital frequency domain multiplexing readout electronics for the next
  generation of millimeter telescopes},'' in {\em Millimeter, Submillimeter,
  and Far-Infrared Detectors and Instrumentation for Astronomy VII},  {\em
  Proc. SPIE} {\bf 9153}, p.~91531A, July 2014, 1407.3161.

\bibitem{Zmuidzinas12}
J.~Zmuidzinas, ``Superconducting microresonators: Physics and applications,''
  {\em Annual Review of Condensed Matter Physics}~{\bf 3}(1), pp.~169--214,
  2012.

\bibitem{Day03}
P.~K. {Day}, H.~G. {LeDuc}, B.~A. {Mazin}, A.~{Vayonakis}, and J.~{Zmuidzinas},
  ``{A broadband superconducting detector suitable for use in large arrays},''
  {\em Nature}~{\bf 425}, pp.~817--821, Oct. 2003.

\bibitem{Irwin2004}
K.~D. Irwin and K.~W. Lehnert, ``{Microwave SQUID multiplexer},'' {\em Applied
  Physics Letters}~{\bf 85}(11), pp.~2107--2109, 2004.

\bibitem{Lehnert07}
K.~W. {Lehnert}, K.~D. {Irwin}, M.~A. {Castellanos-Beltran}, J.~A.~B. {Mates},
  and L.~R. {Vale}, ``{Evaluation of a Microwave SQUID Multiplexer
  Prototype},'' {\em IEEE Transactions on Applied Superconductivity}~{\bf 17},
  pp.~705--709, June 2007.

\bibitem{Mates11PhD}
J.~A.~B. {Mates}, {\em {The Microwave SQUID Multiplexer}}.
\newblock PhD thesis, University of Colorado at Boulder, Dec. 2011.

\bibitem{vanRantwijk16}
J.~{van Rantwijk}, M.~{Grim}, D.~{van Loon}, S.~{Yates}, A.~{Baryshev}, and
  J.~{Baselmans}, ``{Multiplexed Readout for 1000-Pixel Arrays of Microwave
  Kinetic Inductance Detectors},'' {\em IEEE Transactions on Microwave Theory
  Techniques}~{\bf 64}, pp.~1876--1883, June 2016, 1507.04151.

\bibitem{Mates17}
J.~A.~B. {Mates}, D.~T. {Becker}, D.~A. {Bennett}, B.~J. {Dober}, J.~D. {Gard},
  J.~P. {Hays-Wehle}, J.~W. {Fowler}, G.~C. {Hilton}, C.~D. {Reintsema}, D.~R.
  {Schmidt}, D.~S. {Swetz}, L.~R. {Vale}, and J.~N. {Ullom}, ``{Simultaneous
  readout of 128 X-ray and gamma-ray transition-edge microcalorimeters using
  microwave SQUID multiplexing},'' {\em Applied Physics Letters}~{\bf 111},
  p.~062601, Aug. 2017.

\bibitem{Stanchfield16}
S.~M. {Stanchfield}, P.~A.~R. {Ade}, J.~{Aguirre}, J.~A. {Brevik}, H.~M. {Cho},
  R.~{Datta}, M.~J. {Devlin}, S.~R. {Dicker}, B.~{Dober}, D.~{Egan}, P.~{Ford},
  G.~{Hilton}, J.~{Hubmayr}, K.~D. {Irwin}, P.~{Marganian}, B.~S. {Mason},
  J.~A.~B. {Mates}, J.~{McMahon}, M.~{Mello}, T.~{Mroczkowski}, C.~{Romero},
  C.~{Tucker}, L.~{Vale}, S.~{White}, M.~{Whitehead}, and A.~H. {Young},
  ``{Development of a Microwave SQUID-Multiplexed TES Array for MUSTANG-2},''
  {\em Journal of Low Temperature Physics}~{\bf 184}, pp.~460--465, July 2016.

\bibitem{McHugh12}
S.~{McHugh}, B.~A. {Mazin}, B.~{Serfass}, S.~{Meeker}, K.~{O'Brien}, R.~{Duan},
  R.~{Raffanti}, and D.~{Werthimer}, ``{A readout for large arrays of microwave
  kinetic inductance detectors},'' {\em Review of Scientific Instruments}~{\bf
  83}, pp.~044702--044702, Apr. 2012, 1203.5861.

\bibitem{Duan10}
R.~{Duan}, S.~{McHugh}, B.~{Serfass}, B.~A. {Mazin}, A.~{Merrill}, S.~R.
  {Golwala}, T.~P. {Downes}, N.~G. {Czakon}, P.~K. {Day}, J.~{Gao}, J.~{Glenn},
  M.~I. {Hollister}, H.~G. {Leduc}, P.~R. {Maloney}, O.~{Noroozian}, H.~T.
  {Nguyen}, J.~{Sayers}, J.~A. {Schlaerth}, S.~{Siegel}, J.~E. {Vaillancourt},
  A.~{Vayonakis}, P.~R. {Wilson}, and J.~{Zmuidzinas}, ``{An open-source
  readout for MKIDs},'' in {\em Millimeter, Submillimeter, and Far-Infrared
  Detectors and Instrumentation for Astronomy V},  {\em Proc. SPIE} {\bf 7741},
  p.~77411V, July 2010.

\bibitem{Hickish16}
J.~{Hickish}, Z.~{Abdurashidova}, Z.~{Ali}, K.~D. {Buch}, S.~C. {Chaudhari},
  H.~{Chen}, M.~{Dexter}, R.~S. {Domagalski}, J.~{Ford}, G.~{Foster},
  D.~{George}, J.~{Greenberg}, L.~{Greenhill}, A.~{Isaacson}, H.~{Jiang},
  G.~{Jones}, F.~{Kapp}, H.~{Kriel}, R.~{Lacasse}, A.~{Lutomirski},
  D.~{MacMahon}, J.~{Manley}, A.~{Martens}, R.~{McCullough}, M.~V. {Muley},
  W.~{New}, A.~{Parsons}, D.~C. {Price}, R.~A. {Primiani}, J.~{Ray},
  A.~{Siemion}, V.~{van Tonder}, L.~{Vertatschitsch}, M.~{Wagner},
  J.~{Weintroub}, and D.~{Werthimer}, ``{A Decade of Developing Radio-Astronomy
  Instrumentation using CASPER Open-Source Technology},'' {\em Journal of
  Astronomical Instrumentation}~{\bf 5}, pp.~1641001--12, Mar. 2016,
  1611.01826.

\bibitem{Madden17}
T.~J. Madden, T.~W. Cecil, L.~M. Gades, O.~Quaranta, D.~Yan, A.~Miceli, D.~T.
  Becker, D.~A. Bennett, J.~P. Hays-Wehle, G.~C. Hilton, J.~D. Gard, J.~A.~B.
  Mates, C.~D. Reintsema, D.~R. Schmidt, D.~S. Swetz, L.~R. Vale, and J.~N.
  Ullom, ``{Development of ROACH Firmware for Microwave Multiplexed X-Ray TES
  Microcalorimeters},'' {\em IEEE Transactions on Applied
  Superconductivity}~{\bf 27}, pp.~1--4, June 2017.

\bibitem{Gard2018}
J.~D. Gard, D.~T. Becker, D.~A. Bennett, J.~W. Fowler, G.~C. Hilton, J.~A.~B.
  Mates, C.~D. Reintsema, D.~R. Schmidt, D.~S. Swetz, and J.~N. Ullom, ``{A
  Scalable Readout for Microwave SQUID Multiplexing of Transition-Edge
  Sensors},'' {\em Journal of Low Temperature Physics} , Jul 2018.

\bibitem{DoberSPIE18}
B.~Dober {\em et~al.}, ``{Readout demonstration of 512 TES bolometers using a
  single microwave SQUID multiplexer},'' 2018.
\newblock in preparation.

\bibitem{Pozar11}
D.~M. Pozar, {\em Microwave Engineering}, John Wiley \& Sons, Inc., Hoboken,
  NJ, fourth~ed., 2011.

\bibitem{Mates12}
J.~A.~B. {Mates}, K.~D. {Irwin}, L.~R. {Vale}, G.~C. {Hilton}, J.~{Gao}, and
  K.~W. {Lehnert}, ``{Flux-Ramp Modulation for SQUID Multiplexing},'' {\em
  Journal of Low Temperature Physics}~{\bf 167}, pp.~707--712, June 2012.

\bibitem{Kernasovskiy18}
S.~A. {Kernasovskiy}, S.~E. {Kuenstner}, E.~{Karpel}, Z.~{Ahmed}, D.~D. {Van
  Winkle}, S.~{Smith}, J.~{Dusatko}, J.~C. {Frisch}, S.~{Chaudhuri}, H.~M.
  {Cho}, B.~J. {Dober}, S.~W. {Henderson}, G.~C. {Hilton}, J.~{Hubmayr}, K.~D.
  {Irwin}, C.~L. {Kuo}, D.~{Li}, J.~A.~B. {Mates}, M.~{Nasr}, S.~{Tantawi},
  J.~{Ullom}, L.~{Vale}, and B.~{Young}, ``{SLAC Microresonator Radio Frequency
  (SMuRF) Electronics for Read Out of Frequency-Division-Multiplexed Cryogenic
  Sensors},'' {\em Journal of Low Temperature Physics} , May 2018, 1805.08363.

\bibitem{Giachero16}
A.~{Giachero}, D.~{Becker}, D.~A. {Bennett}, M.~{Faverzani}, E.~{Ferri}, J.~W.
  {Fowler}, J.~D. {Gard}, J.~P. {Hays-Wehle}, G.~C. {Hilton}, M.~{Maino},
  J.~A.~B. {Mates}, A.~{Puiu}, A.~{Nucciotti}, C.~D. {Reintsema}, D.~S.
  {Swetz}, J.~N. {Ullom}, and L.~R. {Vale}, ``{Development of
  microwave-multiplexed superconductive detectors for the HOLMES experiment},''
  in {\em Journal of Physics Conference Series},  {\em Journal of Physics
  Conference Series} {\bf 718}, p.~062020, May 2016, 1601.03970.

\bibitem{Till18}
T.~Straumann, R.~Claus, J.~D'Ewart, J.~C. Frisch, G.~Haller, R.~Herbst,
  B.~Hong, U.~Legat, L.~Ma, J.~J. Olsen, B.~A. Reese, L.~Ruckman,
  L.~Sapozhnikov, S.~Smith, J.~A. V\'{a}squez, M.~Weaver, E.~Williams, C.~Xu,
  and A.~Young, ``{The SLAC Common-Platform Firmware for High-Performance
  Systems},'' in {\em {Proceedings, 16th International Conference on
  Accelerator and Large Experimental Physics Control Systems (ICALEPCS 2017):
  Barcelona, Spain, October 8-13, 2017}},  p.~THMPL08, 2018.

\bibitem{ATCAspec}
``{PICMG\textregistered 3.0 Revision 3.0 AdvancedTCA\textregistered Base
  Specification}.''
  \url{https://www.picmg.org/wp-content/uploads/PICMG_ATCA_3_0R3_0.pdf}.
\newblock March 2018.

\bibitem{Frisch17}
J.~Frisch, R.~Claus, J.~D'Ewart, G.~Haller, R.~Herbst, B.~Hong, U.~Legat,
  L.~Ma, J.~Olsen, B.~Reese, L.~Ruckman, L.~Sapozhnikov, S.~Smith,
  T.~Straumann, D.~Van~Winkle, J.~V\'{a}squez, M.~Weaver, E.~Williams, C.~Xu,
  and A.~Young, ``{A FPGA Based Common Platform for LCLS2 Beam Diagnostics and
  Controls},'' in {\em {Proceedings, 5th International Beam Instrumentation
  Conference (IBIC 2016): Barcelona, Spain, September 11-15, 2016}},
  p.~WEPG15, 2017.

\bibitem{Gao07}
J.~{Gao}, J.~{Zmuidzinas}, B.~A. {Mazin}, H.~G. {LeDuc}, and P.~K. {Day},
  ``{Noise properties of superconducting coplanar waveguide microwave
  resonators},'' {\em Applied Physics Letters}~{\bf 90}, p.~102507, Mar. 2007,
  cond-mat/0609614.

\bibitem{RSSI}
T.~Bova and T.~Krivoruchka, ``{Reliable UDP Protocol}.'' DRAFT,
  \url{https://tools.ietf.org/html/ draft-ietf-sigtran-reliable-udp-00}.
\newblock February 1999.

\bibitem{Bennett15}
D.~A. {Bennett}, J.~A.~B. {Mates}, J.~D. {Gard}, A.~S. {Hoover}, M.~W. {Rabin},
  C.~D. {Reintsema}, D.~R. {Schmidt}, L.~R. {Vale}, and J.~N. {Ullom},
  ``{Integration of TES Microcalorimeters With Microwave SQUID Multiplexed
  Readout},'' {\em IEEE Transactions on Applied Superconductivity}~{\bf 25},
  p.~2381878, June 2015.

\bibitem{Choi18}
S.~K. {Choi}, J.~{Austermann}, J.~A. {Beall}, K.~T. {Crowley}, R.~{Datta},
  S.~M. {Duff}, P.~A. {Gallardo}, S.~P. {Ho}, J.~{Hubmayr}, B.~J. {Koopman},
  Y.~{Li}, F.~{Nati}, M.~D. {Niemack}, L.~A. {Page}, M.~{Salatino}, S.~M.
  {Simon}, S.~T. {Staggs}, J.~{Stevens}, J.~{Ullom}, and E.~J. {Wollack},
  ``{Characterization of the Mid-Frequency Arrays for Advanced ACTPol},'' {\em
  Journal of Low Temperature Physics} , June 2018, 1711.04841.

\bibitem{Galitzki18}
N.~Galitzki, A.~Ali, K.~Arnold, P.~C. Ashton, J.~E. Austermann, J.~A. Beall,
  S.~Bryan, P.~G. Calisse, Y.~Chinone, G.~Coppi, K.~D. Crowley, K.~T. Crowley,
  A.~Cukierman, M.~J. Devlin, S.~Dicker, B.~Dober, S.~M. Duff, J.~Dunkley,
  G.~Fabbian, P.~A. Gallardo, M.~Gerbino, N.~Goeckner-Wald, J.~E. Gudmundsson,
  S.~W. Henderson, C.~A. Hill, G.~C. Hilton, S.~P. Ho, L.~A. Howe, J.~Hubmayr,
  B.~Keating, B.~J. Koopman, A.~Kusaka, A.~T. Lee, M.~Limon, F.~Matsuda, P.~D.
  Mauskopf, A.~J. May, J.~McMahon, F.~Nati, M.~D. Niemack, J.~L.
  Orlowski-Scherer, L.~Piccirillo, M.~S. Rao, M.~Salatino, J.~S. Seibert,
  M.~Silva-Feaver, S.~M. Simon, S.~T. Staggs, J.~R. Stevens, A.~Suzuki,
  G.~Teply, R.~Thornton, C.~Tsai, J.~N. Ullom, E.~M. Vavagiakis, M.~R. Vissers,
  E.~J. Wollack, Z.~Xu, and N.~Zhu, ``{The Simons Observatory Cryogenic
  Cameras},'' in {\em Millimeter, Submillimeter, and Far-Infrared Detectors and
  Instrumentation for Astronomy IX},  {\em Proc. SPIE}, 2018.
\newblock In preparation.

\bibitem{CCATprimeSPIE18}
G.~J. {Stacey}, M.~A.~K. {Basu}, N.~{Battaglia}, B.~{Beringue}, F.~{Bertoldi},
  J.~R. {Bond}, P.~{Breysse}, R.~{Bustos}, S.~{Chapman}, D.~T. {Chung},
  N.~{Cothard}, J.~{Erler}, M.~{Fich}, S.~{Foreman}, P.~{Gallardo},
  R.~{Giovanelli}, U.~U. {Graf}, M.~P. {Haynes}, R.~{Herrera-Camus}, T.~L.
  {Herter}, R.~{Hlo{\v z}ek}, D.~{Johnstone}, L.~{Keating}, B.~{Magnelli},
  D.~{Meerburg}, J.~{Meyers}, N.~{Murray}, M.~{Niemack}, T.~{Nikola},
  M.~{Nolta}, S.~C. {Parshley}, D.~{Riechers}, P.~{Schilke}, D.~{Scott},
  G.~{Stein}, J.~{Stevens}, J.~{Stutzki}, E.~M. {Vavagiakis}, and M.~P.
  {Viero}, ``{CCAT-prime: Science with an Ultra-widefield Submillimeter
  Observatory at Cerro Chajnantor},'' {\em ArXiv e-prints} , July 2018,
  1807.04354.

\bibitem{AliCPT18}
H.~Li, S.-Y. Li, Y.~Liu, Y.-P. Li, Y.~Cai, M.~Li, G.-B. Zhao, C.-Z. Liu, Z.-W.
  Li, H.~Xu, D.~Wu, Y.-J. Zhang, Z.-H. Fan, Y.-Q. Yao, C.-L. Kuo, F.-J. Lu, and
  X.~Zhang, ``{Probing Primordial Gravitational Waves: Ali CMB Polarization
  Telescope},'' {\em National Science Review} , p.~nwy019, 2018, 1710.03047.

\bibitem{Li18}
D.~Li, B.~K. Alpert, D.~T. Becker, D.~A. Bennett, G.~A. Carini, H.-M. Cho,
  W.~B. Doriese, J.~E. Dusatko, J.~W. Fowler, J.~C. Frisch, J.~D. Gard,
  S.~Guillet, G.~C. Hilton, M.~R. Holmes, K.~D. Irwin, V.~Kotsubo, S.-J. Lee,
  J.~A.~B. Mates, K.~M. Morgan, K.~Nakahara, C.~G. Pappas, C.~D. Reintsema,
  D.~R. Schmidt, S.~R. Smith, D.~S. Swetz, J.~B. Thayer, C.~J. Titus, J.~N.
  Ullom, L.~R. Vale, D.~D. Van~Winkle, A.~Wessels, and L.~Zhang, ``{TES X-ray
  Spectrometer at SLAC LCLS-II},'' {\em Journal of Low Temperature Physics} ,
  2018.
\newblock In preparation.

\end{thebibliography}
\bibliographystyle{spiebibweprint} 
\end{document}